\documentclass[a4paper,fleqn,usenatbib]{mnras}

\usepackage[T1]{fontenc}
\usepackage{ae,aecompl}

\usepackage{graphicx}	% Including figure files
\usepackage{amsmath}	% Advanced maths commands
\usepackage{amssymb}	% Extra maths symbols

\title[LMC old globular clusters]{Evidence of differential tidal effects in the
old globular cluster population of the  Large
Magellanic Cloud}

\author[A.E. Piatti et al.]{
A.E. Piatti$^{1,2}$\thanks{E-mail: andres@oac.unc.edu.ar}
and A.D. Mackey$^3$
\\
$^{1}$Consejo Nacional de Investigaciones Cient\'{\i}ficas y T\'ecnicas, Av. Rivadavia 1917, 
C1033AAJ, Buenos Aires, Argentina\\
$^{2}$Observatorio Astron\'omico, Universidad Nacional de C\'ordoba, Laprida 854, 5000, 
C\'ordoba, Argentina\\
$^3$ Research School of Astronomy \& Astrophysics, Australian National University, 
Canberra, ACT 2611, Australia\\
}

\date{Accepted XXX. Received YYY; in original form ZZZ}

\pubyear{2017}

\begin{document}
\label{firstpage}
\pagerange{\pageref{firstpage}--\pageref{lastpage}}
\maketitle

% Abstract of the paper
\begin{abstract}

We present for the first time extended stellar density and/or surface brightness radial profiles for almost all the known 
Large Magellanic Cloud (LMC) old globular clusters (GCs).
These were built from DECam images and reach out to $\sim$ 4
times the GCs' tidal radii. 
The background subtracted radial profiles
reveal that the GCs located closer than $\sim$ 5 kpc from the LMC centre
contain an excess of stars in their outermost regions with respect
to the stellar density expected from a King profile. Such a residual amount of stars
- not seen in GCs located farther than $\sim$ 5 kpc from the LMC centre-,
as well as the GCs' dimensions, show a clear dependence with the GCs' positions in the 
galaxy, in the sense that, the farther the GC from the centre of the LMC, the larger both 
the excess of stars in its outskirts and size. 
%This trend still holds when
%GCs located between $\sim$ 5 and 15 kpc from the LMC centre are added to the present
%sample. 
Although the masses of GCs located inside and outside $\sim$ 5 kpc are commensurate, 
the outermost regions of GCs located closer than $\sim$ 5 kpc from
the LMC centre appear to have dynamically evolved more quickly.
These outcomes can be fully interpreted in the light of the known GC radial 
velocity disc-like kinematics, from which GCs have been somehow mostly experiencing the 
influence of the LMC gravitational field at their respective mean distances from the LMC centre.
%GCs placed
%in orbits closer to the core of the galaxy have suffered relatively more severe effects,
%for instance, that more GC stars moved outwards reaching the GC outskirts. Furthermore,
%the gradual decrease of the LMC gravitational field with the distance from its centre
%has also allowed GC to expand more as they occupy farther positions in the galaxy.

\end{abstract}

% Select between one and six entries from the list of approved keywords.
% Don't make up new ones.
\begin{keywords}
techniques: photometric -- galaxies: individual: LMC --
galaxies: star clusters: general 
\end{keywords}

%%%%%%%%%%%%%%%%%%%%%%%%%%%%%%%%%%%%%%%%%%%%%%%%%%

%%%%%%%%%%%%%%%%% BODY OF PAPER %%%%%%%%%%%%%%%%%%

\section{Introduction}

Recently \citet[][hereafter WK17]{wagnerkaiseretal2017} used HST data to show that
metal-poor old globular clusters (GCs) in the inner halo of the Milky Way (MW) and 
in the Large Magellanic Cloud (LMC) are highly synchronized, in the sense that they 
seem to be coeval to 0.2 $\pm$ 0.4 Gyr. Because their masses are also similar
\citep[][hereafter MG03]{sb2017,mg2003}, it becomes interesting to
investigate whether such a synchronization has reached other astrophysical 
properties linked to them, such as structural parameters, relaxation times, etc.

Within the Galactic globular cluster (GGC) population, the presence of extra-tidal 
features is frequently seen, either as tidal tails, or extra-tidal stellar populations, 
or extent diffuse halo-like structures 
\citep[e.g.][]{odenetal2003,correntietal2011,carballobelloetal2012,kuzmaetal2017,myeongetal2017,naverreteetal2017,p17c}. 
Nevertheless, as pointed out by \citet{p17d}, a renewed overall study of the external regions of 
GGCs is needed to reliably characterize them, and hence to investigate whether there is any 
connection between detected extra-tidal features with the GGCs' dynamical histories in the Galaxy. 
If ages and metallicities led WK17 to conclude on the synchronicity of GC formation
in the MW and the LMC, the comparison of their structural parameters (e.g. core and tidal radii) could
shed light about any synchronicity of their dynamical histories as a result of their internal dynamics and
tidal interactions with their host galaxies. 

MG03 derived accurate \citet{king62}'s core radii ($r_c$) for 
five out of the six LMC GCs analysed by WK17, namely, NGC\,1466, 1841, 2210,
2257 and Hodge\,11. They could not estimate tidal radii ($r_t$) because of the limited field-of-view
of the HST camera. The sixth GC in the WK17 sample, Reticulum, has $r_c$ and $r_t$ values
estimated from photographic plates by \citet{pk1977}. As far as we are aware, there is no
other study on the external regions of these LMC GCs located farther than $\sim$ 5 kpc from the LMC centre. For this reason, we took advantage of
DECam images \citep[3 square degree field-of-view, pixel size=0.263$\arcsec$;][]{flaugheretal2015} 
to perform a  sound analysis of their outskirts and to compare their structural properties with those of GGCs.

The LMC harbours an additional nine GCs located inside $\sim$ 5 kpc from its centre,
namely: NGC\,1754, 1786, 1835, 1898, 1916, 1928, 1939, 2005 and 2019. Because of their
closer proximity to the LMC main body, they could have experienced a different tidal influence 
from their host galaxy, resulting in a distinctive behaviour of their structural properties.
Similarly to the above subsample of LMC GCs, we did not find any other study  on their external 
regions, so that we added them to this study by constructing accurate 
surface brightness radial profiles out to $\sim$ 4 times their tidal radii, using 
DECam images as well.

The paper is organized as follows: Section\,2 introduces the data sets used and describes their processing.
In Section\,3 we explain how we constructed the stellar density and surface brightness radial profiles. 
It also presents estimates of a wide variety of dynamical and structural parameters from
the constructed profiles that results in an homogeneous compilation for almost all
ancient GCs in the LMC. Section\,4 deals with the analysis of the resultant structural 
parameters and compares them with those of the GGCs, in order to establish any connection between them, 
as WK17 did for their ages and metallicities. While searching for a possible relationship between these dynamical 
and structural properties, we discovered that the existence of extra-tidal structures 
is dependent on the distance of a cluster from the LMC centre: those
GCs closer than $\sim$ 5 kpc to the galaxy centre do exhibit such extended features. 
We summarise our results in Section 5.

\section{Data handling}

\subsection{Resolved GCs}

By searching the National Optical Astronomy Observatory 
(NOAO) Science Data Management (SDM) Archives\footnote{http://www.noao.edu/sdm/archives.php.}
we found DECam images taken in the field of NGC\,1841, 2210 and Hodge\,11 by the
Survey of the Magellanic Stellar History \citep[SMASH;][]{nideveretal2017}, and in the field 
of Reticulum by the Dark Energy Survey \citep[DES;][]{abbottetal2016}.
The images were taken using the $ugriz$ filters, which are similar, but not identical to, the comparable SDSS filters. We downloaded the deepest publicly available images containing the aforementioned GCs,
discarding those in which the clusters fall in gaps or are close to their margins. 
We preferred resampled images, because they have been corrected for distortion and have better
astrometric coordinate solutions \citep{valdesetal2014}. They consist of 120 s co-added $i$ 
images (90 s for Reticulum) with typical FWHMs of 0.95$\arcsec$.

\begin{figure*}
\includegraphics[width=\columnwidth]{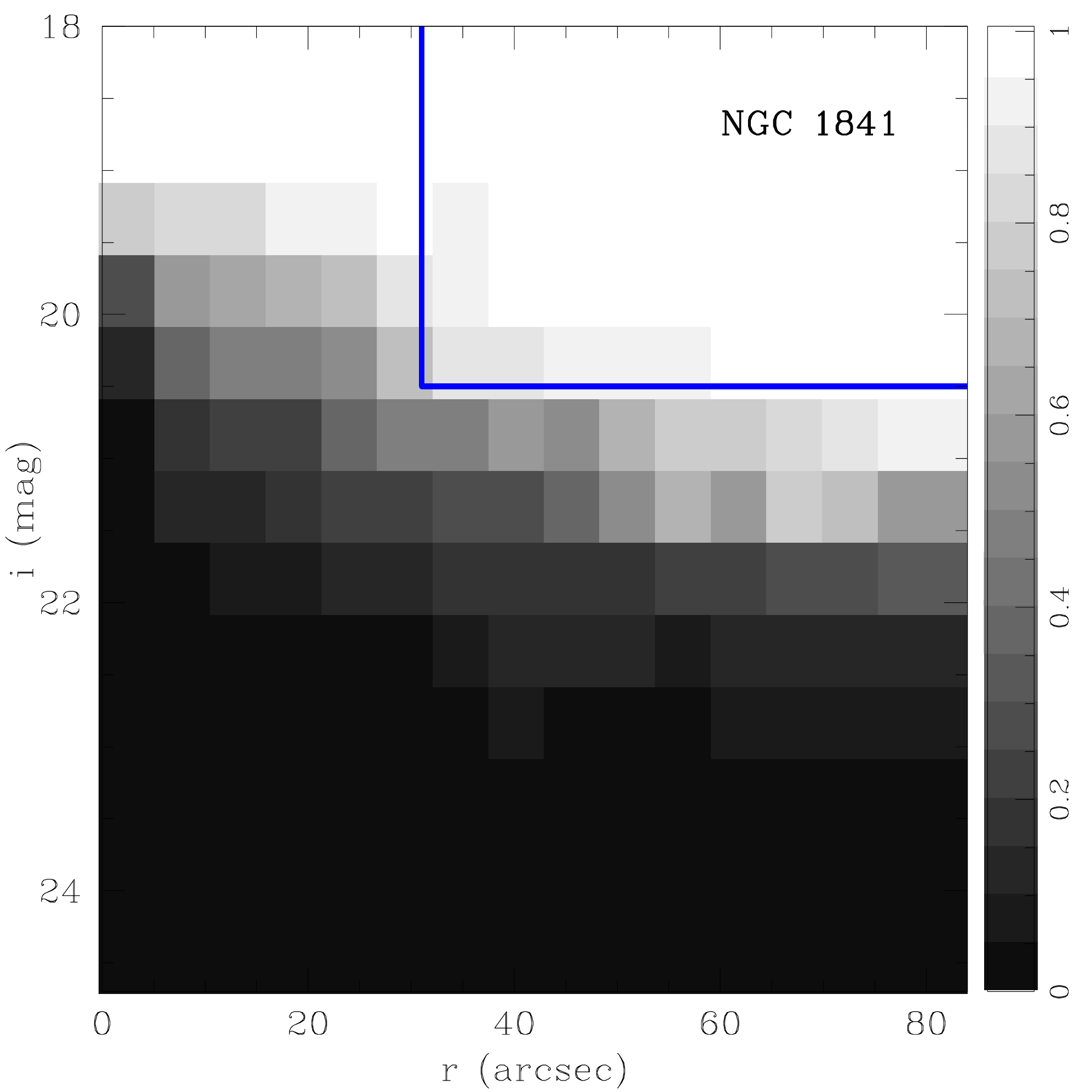}
\includegraphics[width=\columnwidth]{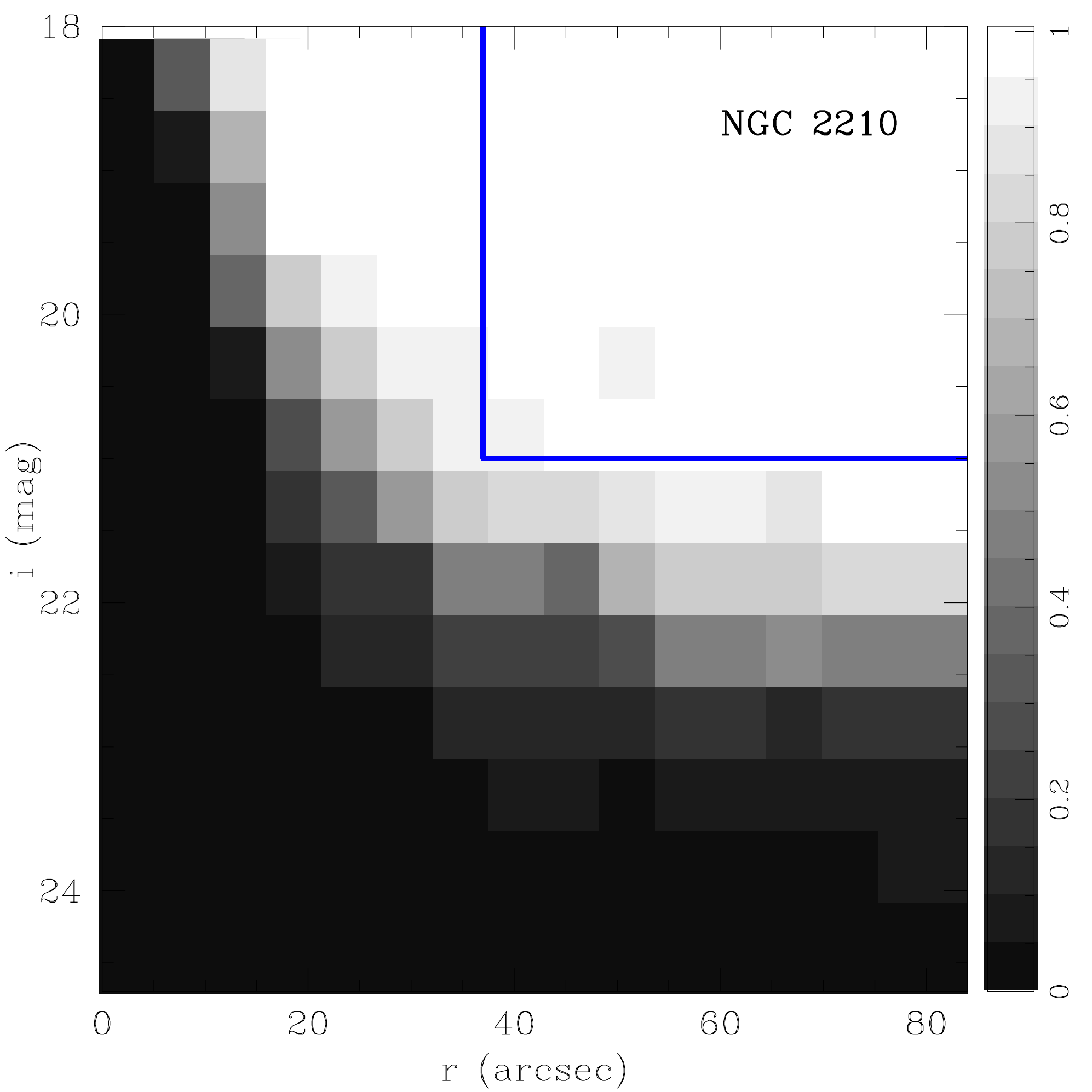}
\includegraphics[width=\columnwidth]{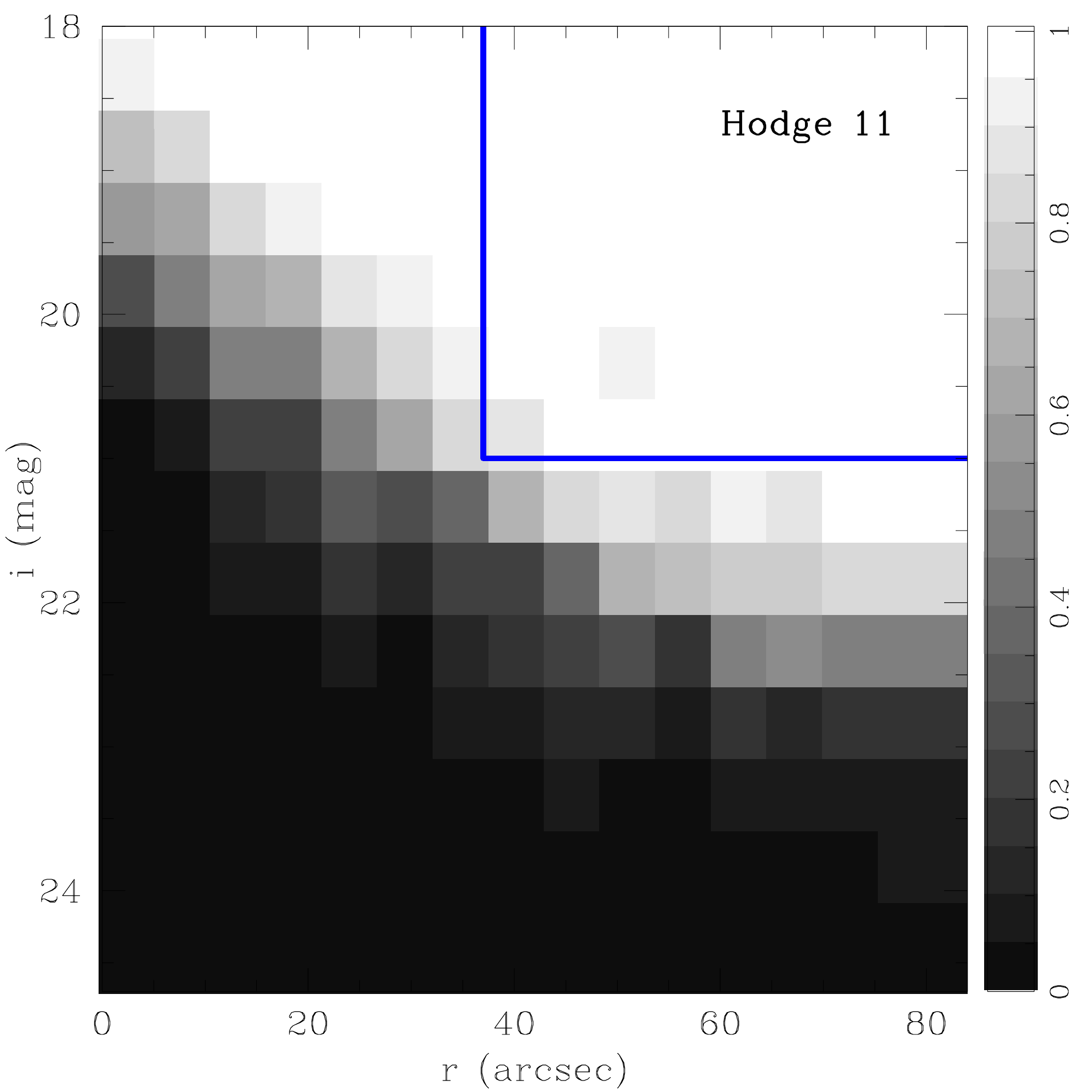}
\includegraphics[width=\columnwidth]{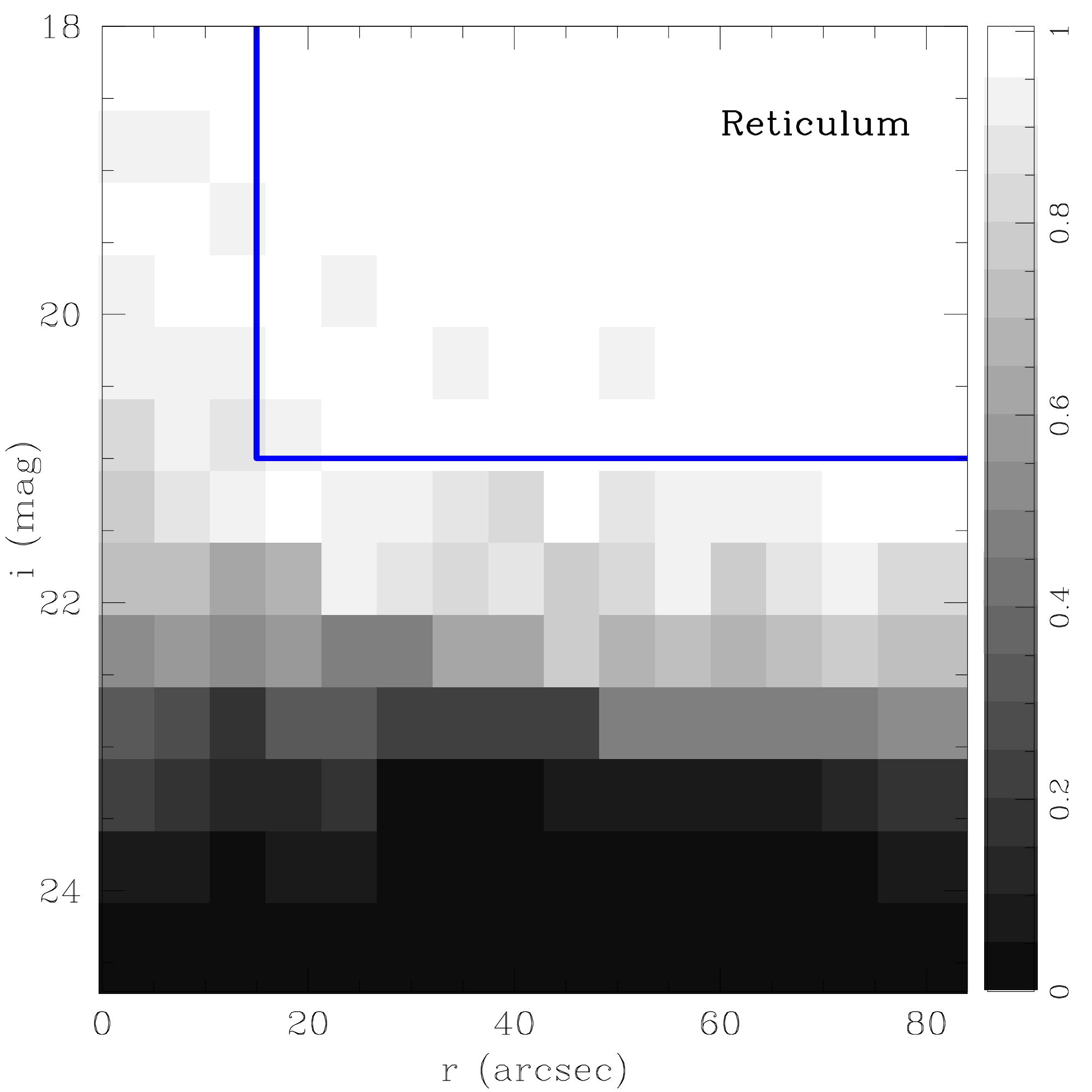}
    \caption{Variation of the photometric completeness as a function of the instrumental 
$i$ magnitude and the distance from the cluster centre, according to the grey-scale
bar placed to the right margin of each panel. Blue lines delineate the region of stars used to
build the stellar density profiles. 
}
   \label{fig:fig1}
\end{figure*}

We obtained point-spread-function (PSF) photometry using the stand-alone DAOPHOT package 
\citep{setal90}. A series of tasks comprising star finding and aperture photometry, PSF 
modelling with functions quadratically varying, and the use of the resulting PSFs to obtain 
instrumental magnitudes were performed iteratively three times on previously created 
subtracted images to find and measure magnitudes of additional fainter stars 
\citep[see, e.g.,][]{p17c,p17e,piattietal2017a}. Bona fide stellar objects were
successfully isolated by using roundness values between -0.5 and 0.5 and sharpness values 
between 0.2 and 1.0. 

We also performed extensive artificial star tests around the cluster regions in order to
accurately map the completeness of our photometry in terms of photometric depth and spatial
dependence with the distance from the cluster centre. In doing this we repeated the PSF 
photometry recipes referred above -- including the three passes to measure fainter stars -- 
on a thousand created images per cluster with nearly 5 per cent  added stars distributed
appropriately according to the cluster stellar density profile and covered magnitude range 
\citep[see, e.g.][]{p15,pb16a,pc2017}. The resultant completeness functions are depicted in 
Fig.~\ref{fig:fig1}.

\subsection{Unresolved GCs}

We also made use of DECam images downloaded from the NOAO SDM Archives taken by SMASH 
in the field of NGC\,1754, 1786, 1835, 1898, 1916, 1928, 1939, 2005 and 2019.
From the whole collection of retrieved resampled images, we 
discarded those that have surpassed the detector saturation limits or show proccesing defects (e.g., image damaged by 
unsuccessful CCD reading or charge transfer), in addition to the constrains used
for resolved GCs. We finally selected images taken using the 
$g,i$ filters. All these nine GCs are unresolved extended objects in the DECam images (GC typical size of $\sim$ 
30-80 arcsec in radius), so that we built surface brightness radial profiles from 
integrated
aperture photometry, as described in Section 3.2.

\section{Data analysis}

\subsection{Stellar density radial profiles}

We built stellar density radial profiles for the four resolved LMC GCs.
We first determined the cluster photometric centres by using stars with $i$ magnitudes
at 100 per cent completeness level, from which we applied a kernel density estimator (KDE)
technique. Particularly, we employed the KDE routine within AstroML \citep{astroml},
which has the advantage of not depending on the bin size and starting point to build a
stellar density map, as it is the case whenever star counts and histograms are produced.
The only free parameter is the so-called bandwidth, which refers to the FWHM of the
Gaussians used to build the stellar density map. Bandwidths were varied from 2 up to
10 times the images' FWHMs without noticing any changes in the  photometric centres 
larger than the estimated uncertainties (typical $\sigma$ $\sim$ 1.0$\arcsec$). In the 
case of NGC\,2210 we used the IRAF\footnote{IRAF is distributed by the National 
Optical Astronomy Observatories, which is operated by the Association of 
Universities for Research in Astronomy, Inc., under contract with the National 
Science Foundation.} {\sc n2gaussfit} task instead, because the central cluster region is not
resolved by the DECam images.

\begin{figure*}
\includegraphics[width=\columnwidth]{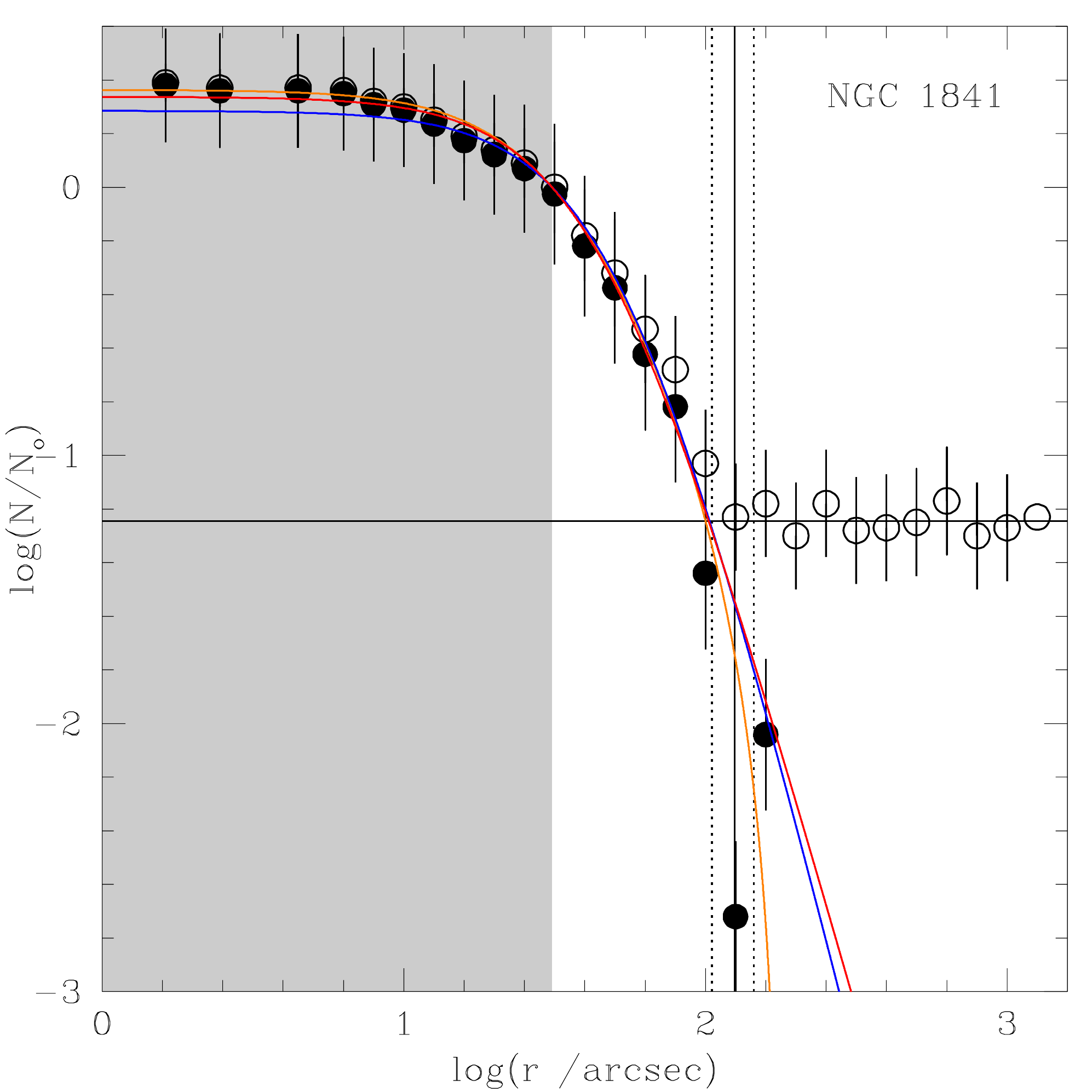}
\includegraphics[width=\columnwidth]{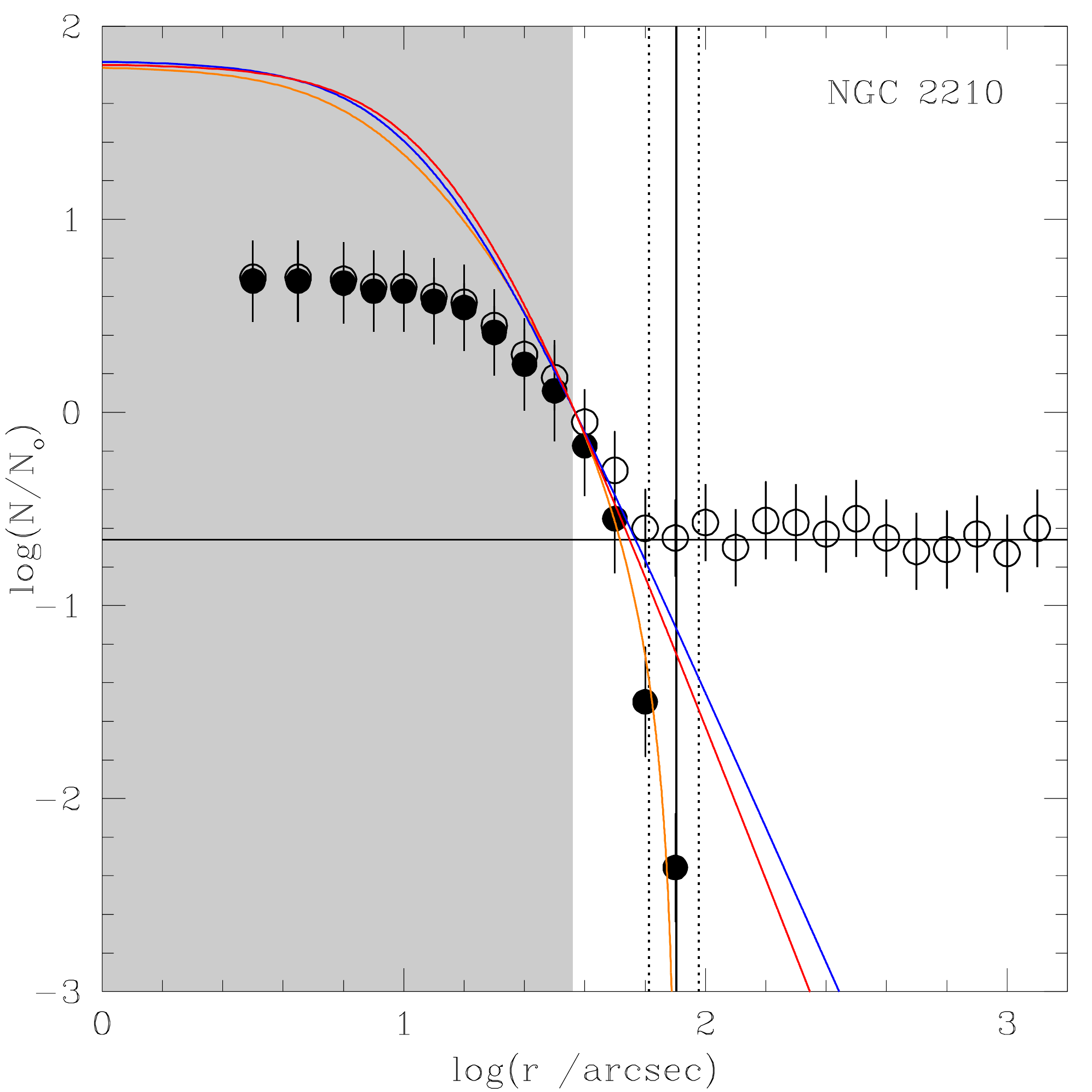}
\includegraphics[width=\columnwidth]{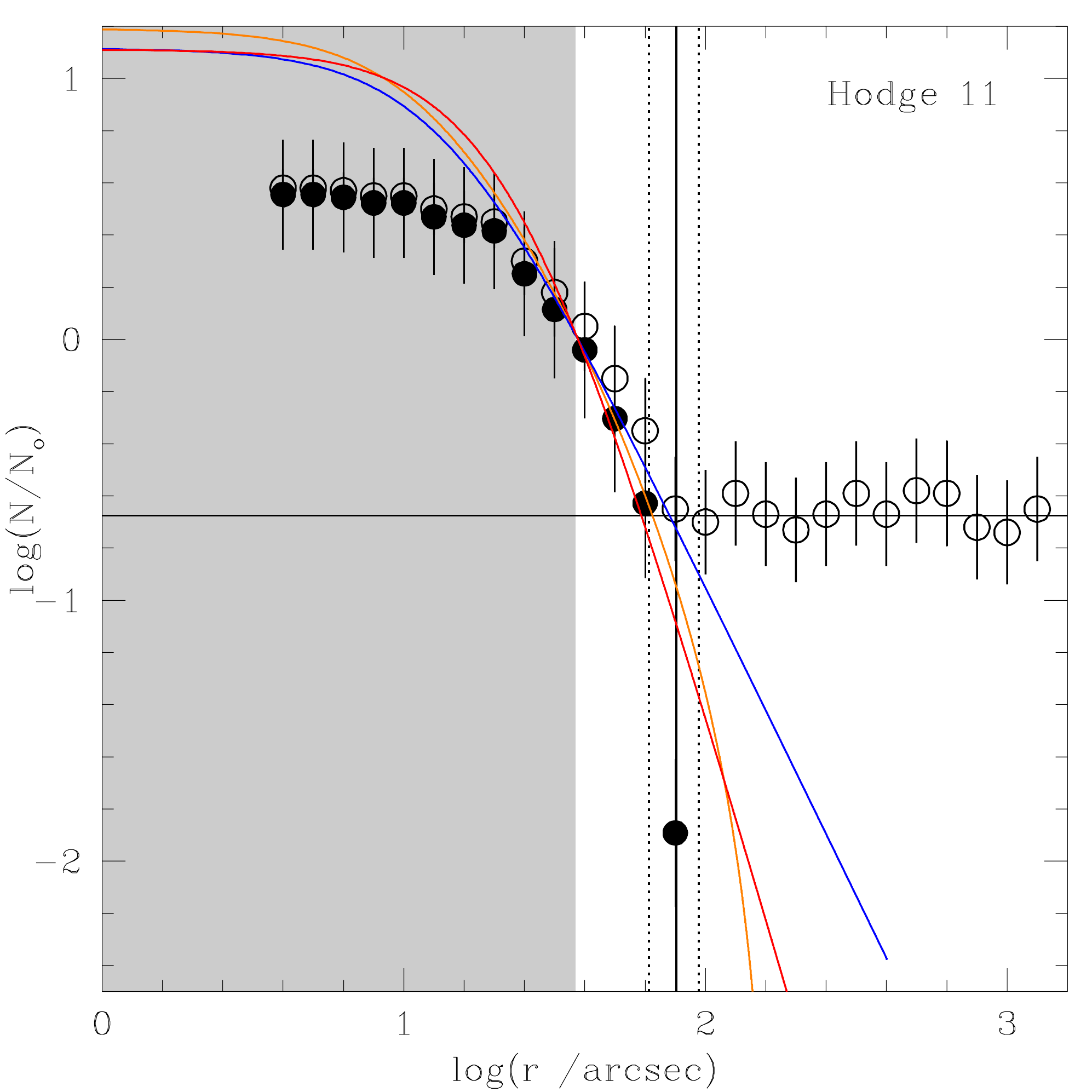}
\includegraphics[width=\columnwidth]{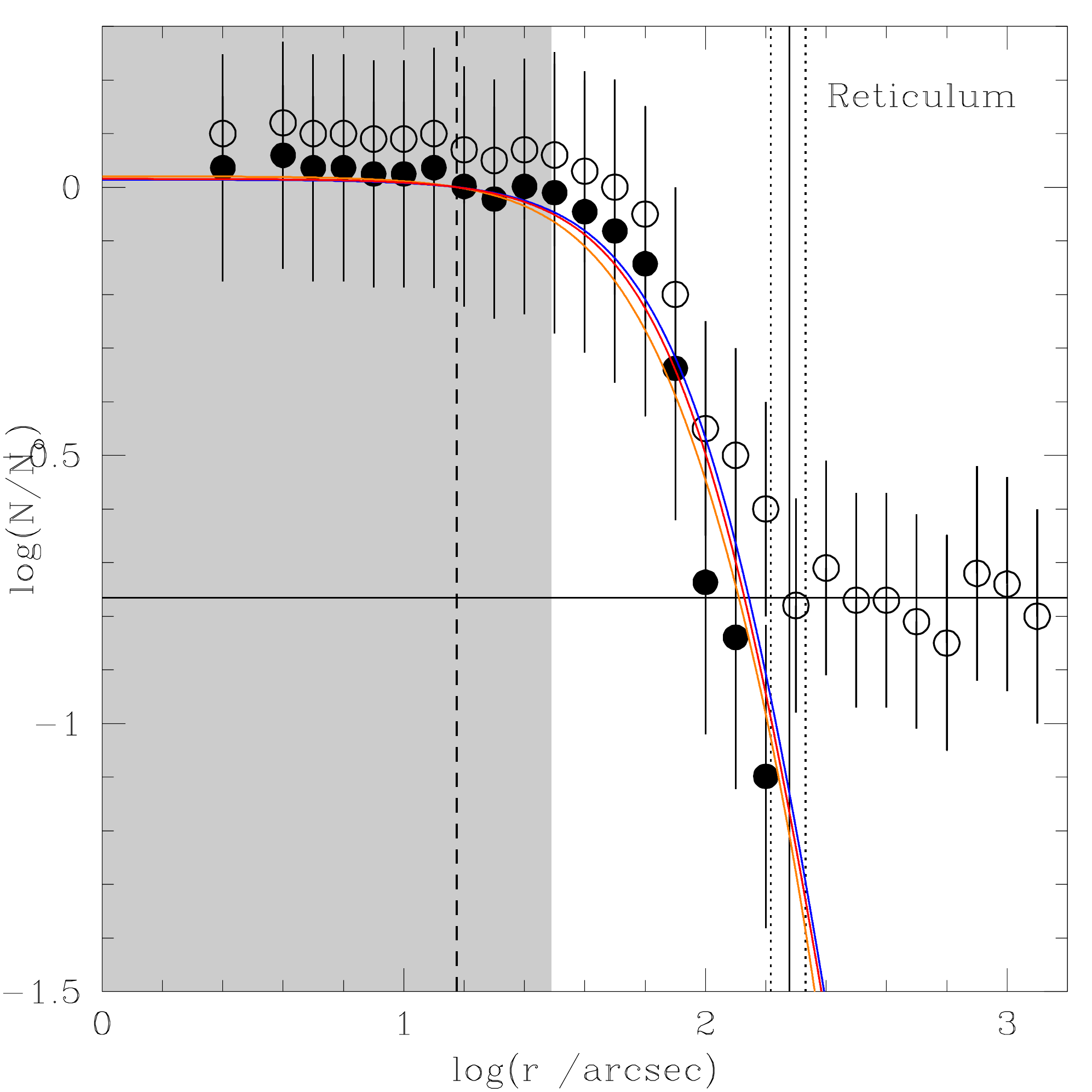}
    \caption{Measured and background subtracted radial stellar density profiles represented with open and 
    filled circles, respectively. The horizontal line represents the mean background level, while the 
    solid and dotted vertical lines represent the derived $r_{cls}$ and its uncertainty.
 The shaded area was not consider to perform the fits of
    \citet{king62}, \citet{eff87} and \citet{plummer11} profiles, which were superimposed with
    orange, blue and red lines, respectively.  The observed mismatch towards the centres is due 
to photometry incompleteness.
    }
   \label{fig:fig2}
\end{figure*}

Stellar density maps were constructed by employing the KDE routine over a subsample
of stars with $i$ magnitudes brighter than those for the 90 per cent completeness level.
Since this is a compromise between desirable photometric depth and distance from the cluster
centre, we chose radii and $i$ magnitude limits that allowed us to reach relatively 
faint stars located reasonably inside the cluster main bodies. Fig.~\ref{fig:fig1} shows
those chosen limits in $i$ magnitude and radius represented by the horizontal and vertical
blue lines, respectively. From the resultant stellar density maps, we built
the cluster stellar density profiles using the above photometric centres and averaging
the generated stellar density values for annular regions of $\Delta$log(r /arcsec)= 0.1
wide. These measured stellar density profiles are shown in Fig.~\ref{fig:fig2} with
open circles with the respective errorbars. From them, the mean background levels were
estimated by averaging those values for log(r /arcsec) $\ge$ 2.6 
(horizontal lines in Fig.~\ref{fig:fig2}) and subtracted from the
measured stellar density profiles. Although clusters have quite different profiles
we chose that value because is readily visible from Fig.~\ref{fig:fig2} 
that a meaningful mean background can be obtained from values outwards that limit.
The background subtracted profiles are depicted
with filled circles in Fig.~\ref{fig:fig2}. In this case, the errorbars come from considering
in quadrature the uncertainties of the measured density profiles and the dispersion
of the background levels. Additionally, from the intersection of the mean background level
and the GC stellar density profile we derived the observed cluster radii ($r_{cls}$). The mean
values and errors are illustrated with solid and dotted vertical lines in Fig.~\ref{fig:fig2}.

MG03 fitted the surface brightness profiles of NGC\,1841, 2210 and Hodge\,11 - obtained 
from HST data that reach $\sim$ 76 arcsec out of the cluster centres - with 
\citet[][hereafter EFF]{eff87}'s models through the expression:

\begin{equation}
\mu(r) = \mu_o \left(1 + \frac{r^2}{a^2}\right)^{-\gamma/2}
\end{equation} 

\noindent where $\mu_o$, $a$ and $\gamma$ are the central surface brightness, a measure of 
the core radius and the the power-law slope at large radii, respectively. We used 
their $a$ and $\gamma$ values to overplot EFF models 
onto the background subtracted stellar density profiles, which very well match the outer 
cluster regions as well, as can be seen in Fig.~\ref{fig:fig2} (blue lines). For Reticulum,
we performed here our own fit ($a$ = 29.1$\pm$1.7 pc, $\gamma$ = 4.2$\pm$0.3) by
using a grid of ($a$, $\gamma$) values to fit its stellar radial profile by $\chi^2$
minimization. Notice that 
the stellar density profiles are affected by photometry completeness less than 90 per cent  
across the shaded areas of Fig.~\ref{fig:fig2}. In the cases of NGC\,2210 and 
Hodge\,11 such an incompleteness is more severe than for NGC\,1841 and Reticulum.

We also used \citet{king62}'s profiles with the $rc$ values derived by MG03 to find the 
$r_t$ ones that best reproduce the stellar density cluster profiles. Because of
the limitation of this method to derive $r_t$ values, we constrained their use in the
subsequent analysis only to estimate concentration parameters $c$ ($\equiv$ log($r_t$/$r_c$)),
allowing relatively larger uncertainties (see Table~\ref{tab:table1}). For Reticulum, we also
derived $r_c$. In order to have
independent estimates of the cluster half-light radii ($r_h$), we fitted \citet{plummer11}'s
models using the
relation $r_h$ $\sim$ 1.3$\times$$a$; $a$ being the only free parameter
to adjust. Both, King and Plummer resultant curves are illustrated in Fig.~\ref{fig:fig2}
with orange and red lines, respectively. From the resulting $r_h$ values and the asymptotic masses
($M_{\infty}$) derived by MG03 (by \citet{setal92} for Reticulum), we calculated half-mass relaxation times 
using the equation of \citet{sh71}

\begin{equation}
t_r = \frac{8.9\times 10^5 M_{\infty}^{1/2} r_h^{3/2}}{\bar{m} log_{10}(0.4M_{\infty}/\bar{m})}
,\end{equation}

\noindent where $\bar{m}$ is the average mass of the cluster stars. For simplicity we assumed a 
constant average stellar mass of 0.75 M$_{\odot}$, which corresponds to a star at the main sequence
turnoff of these LMC GCs \citep{betal12}. Finally, we computed the values of Jacobi radii from the 
expression \citep{cw90}

\begin{equation}
r_J = (\frac{M_{\infty}}{3 M_{LMC}})^{1/3}\times d_{deproj}
,\end{equation}

\noindent where $M_{LMC}$ is the LMC mass contained in a volume of radius equals to the cluster
deprojected galactocentric distance ($d_{deproj}$). The latter were calculated by assuming that the 
GCs are part of a disc having an inclination $i$ = 35.14$\degr$ and a position angle of the line 
of nodes of $\Theta$ = 129.51$\degr$ \citep{betal15}. For $M_{LMC}$ we used three available values
of 0.5, 1.7 and 18 $\times$10$^{10}$ $M_\odot$ contained inside  $d_{deproj}$ = 4, 9 and 20 kpc, respectively 
\citep{beslaetal2012} and derived the fitted equation $M_{LMC}$ = 0.023 $(d_{deproj}$ /kpc)$^3$ ($\times$10$^9$ $M_\odot$), 
which we used to interpolate the respective LMC mass values. Table~\ref{tab:table1} lists all the 
relevant astrophysical properties estimated with their respective uncertainties.

\begin{table*}
%\centering
\caption{Derived properties of resolved LMC GCs.}
\label{tab:table1}
\begin{tabular}{@{}lccccccccc}\hline
ID        &  $d_{deproj}$ & $r_c$$^a$      & $r_h$          &  $r_{cls}$     &  $r_t$          & $r_J$           & $t_r$   & age$^b$   & log($M_{\infty}$ /$M_\odot$)$^c$ \\
          &  (kpc)    & (pc)           & (pc)           & (pc)           & (pc)            & (pc)            & (Gyr)   & (Gyr)  &                 \\\hline  
NGC\,1841 & 15.55 &  7.77$\pm$0.17 & 14.75$\pm$1.26 & 31.51$\pm$5.04 & 50.42$\pm$12.61 & 109.7$\pm$22.69 & 4.8$\pm$1.1   & 13.77$\pm$1.70& 5.12$\pm$0.20\\
NGC\,2210 &  4.37 &  1.99$\pm$0.06 &  4.46$\pm$0.49 & 19.62$\pm$3.68 & 19.62$\pm$4.91  & 119.2$\pm$9.81  & 1.2$\pm$0.2   & 11.63$\pm$1.50& 5.48$\pm$0.10\\
Hodge\,11 &  4.53 &  2.95$\pm$0.16 &  7.65$\pm$0.25 & 20.08$\pm$3.76 & 42.66$\pm$2.51  & 138.5$\pm$32.62 & 2.9$\pm$0.8   & 13.92$\pm$2.10& 5.63$\pm$0.24\\
Reticulum     & 10.62 & 18.82$\pm$1.26 & 36.39$\pm$3.76 & 47.68$\pm$6.27 &112.90$\pm$12.55 & 76.0$\pm$40.15  & 19.0$\pm$13.0 & 13.09$\pm$2.10& 5.15$\pm$0.25\\
\hline
\end{tabular}

\noindent $^a$ Taken from MG03, except Reticulum's value which is derived in this work; 
 $^b$ taken from WK17; $^c$  taken from MG03, and from \citet{setal92} for Reticulum.
Note: to convert 1 arcsec to pc, we use the following expression,10$\times$10$^{(m-M)_o/5}$sin(1/3600)
where $(m-M)_o$ is the true distance modulus taken from WK17.
\end{table*}

\subsection{Surface brightness radial profiles}

We produced surface brightness radial profiles for the nine unresolved LMC GCs, as well as
for the resolved ones for completeness purposes.
We first determined the centres of
gravity by using the {\sc n2gaussfit} task within IRAF. We started with initial guesses 
for the background level, the
cluster centre, the amplitude (brightness above background), the full-width-half-maximum,
the position angle and the ellipticity, and run the routine using square boxes of 10 
(2.63$\arcsec$) up to 40 (10.52$\arcsec$) pixels a side, in steps of 5 (1.32$\arcsec$) 
pixels. Then, we averaged all the output central coordinates with typical resulting 
uncertainties of $\sim$ 0.5 (0.13$\arcsec$) pixels.

We measured instrumental $g,i$ integrated magnitudes around the cluster centres in
concentric circles with radii from 5 (1.32$\arcsec$) up to 1000 (4.38$\arcmin$) pixels,
increasing the circle radius in steps of 5 (1.32$\arcsec$) pixels. Therefore, we obtained
200 different integrated magnitude values per filter across the cluster fields. 
Because of the high S/N ratio (measured flux to background level) photometric errors 
resulted always smaller than 0.005 mag and decreased as the circle radius increased.
In order to obtain  mean integrated magnitudes per unit area as a function of the
distance from the cluster centres, we computed the difference of measured fluxes
between two adjacent circles, then divided the resulting flux by the respective
annular area and computed its integrated magnitude. The errors were calculated using
the expression  2$\times$$\sigma$($mag_j$)/($flux_{j+1}/flux_j$ -1), where
$\sigma$($mag_j$), $flux_j$ and $flux_{j+1}$ represent the photometric error of 
the integrated magnitude for the $j$-th circle, and the measured fluxes
at the $j$ and $j+1$-th circles, respectively. The adopted equation comes from 
error propagation of the difference between two magnitudes, assuming $\sigma$($mag_j$) 
= $\sigma$($mag_{j+1}$). We finally averaged all the available values in intervals
of log($r$ /arcsec) = 0.1. 

\begin{figure}
\includegraphics[width=\columnwidth]{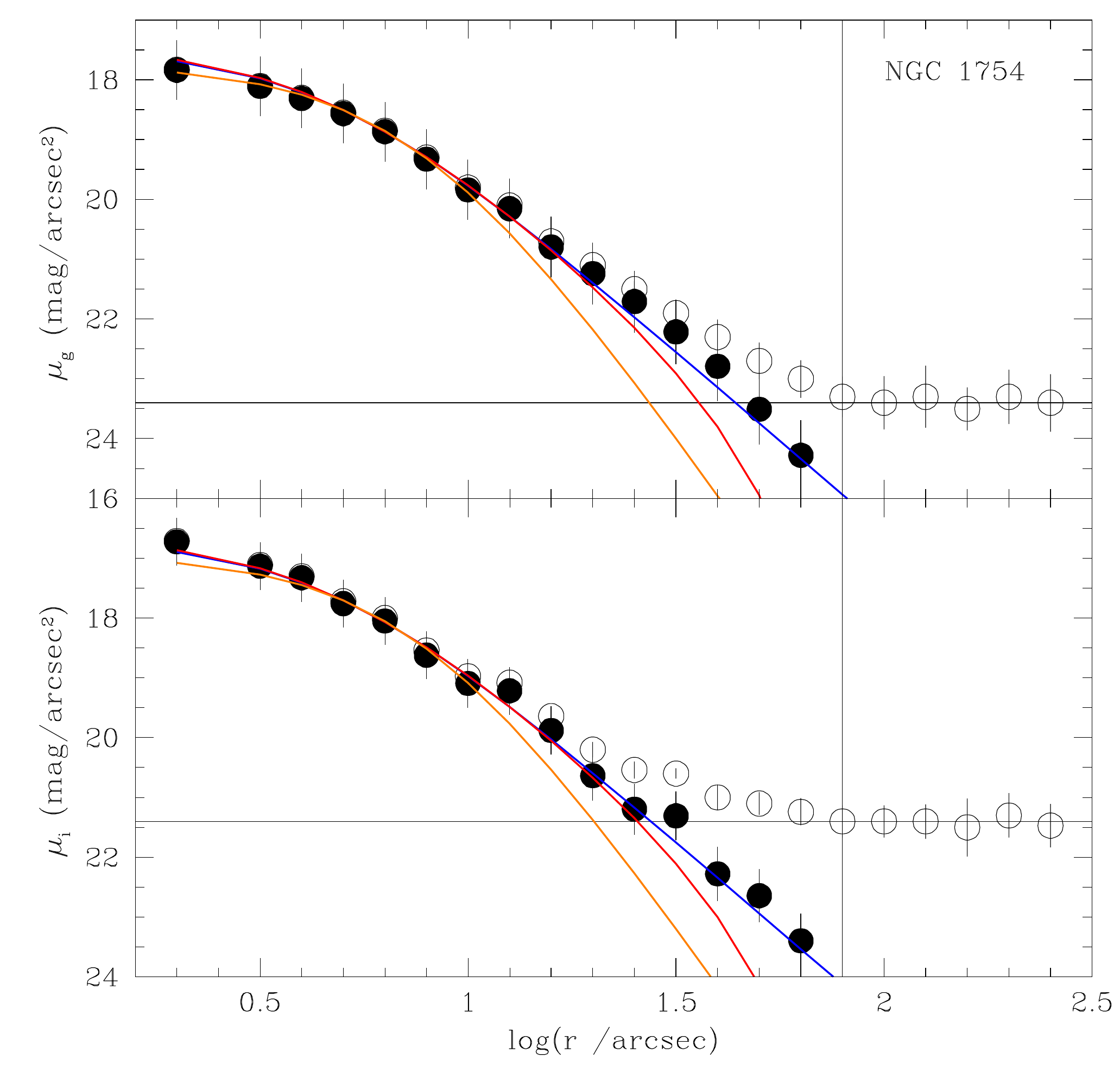}
    \caption{Measured and background subtracted surface brightness radial profiles
of NGC\,1754 represented with open and filled circles, respectively,  for $g$ (top)
and $i$ (bottom) filters.
The corresponding
errobars are also depicted. The mean background level and the position of the
cluster radius are illustrated with horizontal and vertical lines, respectively. Blue,
red and orange lines are the fitted EFF, King and Plummer models, respectively.}
   \label{fig:fig3}
\end{figure}

\begin{figure}
\includegraphics[width=\columnwidth]{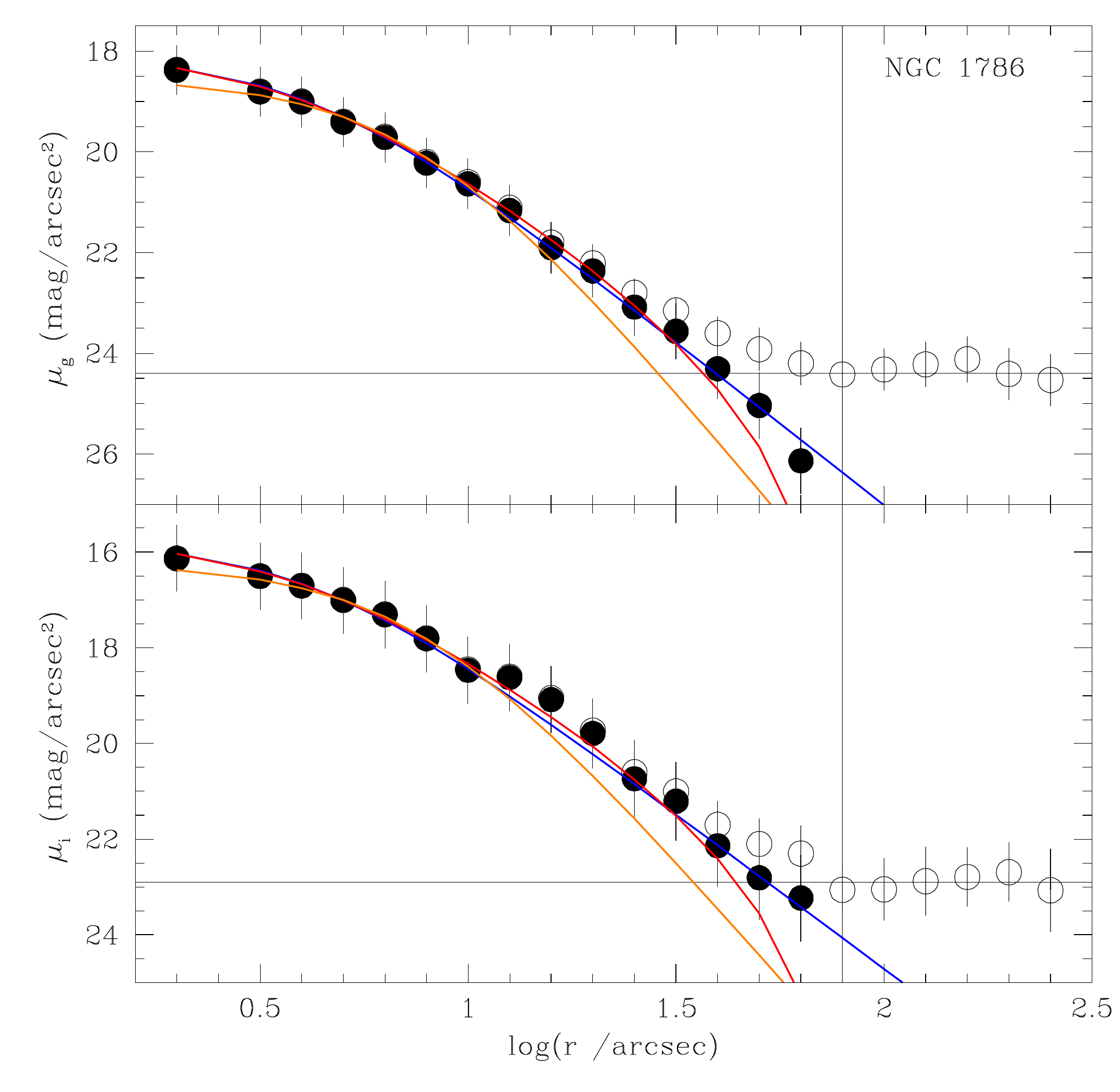}
    \caption{Same as Fig.~\ref{fig:fig3}, for NGC\,1786.}
   \label{fig:fig4}
\end{figure}

\begin{figure}
\includegraphics[width=\columnwidth]{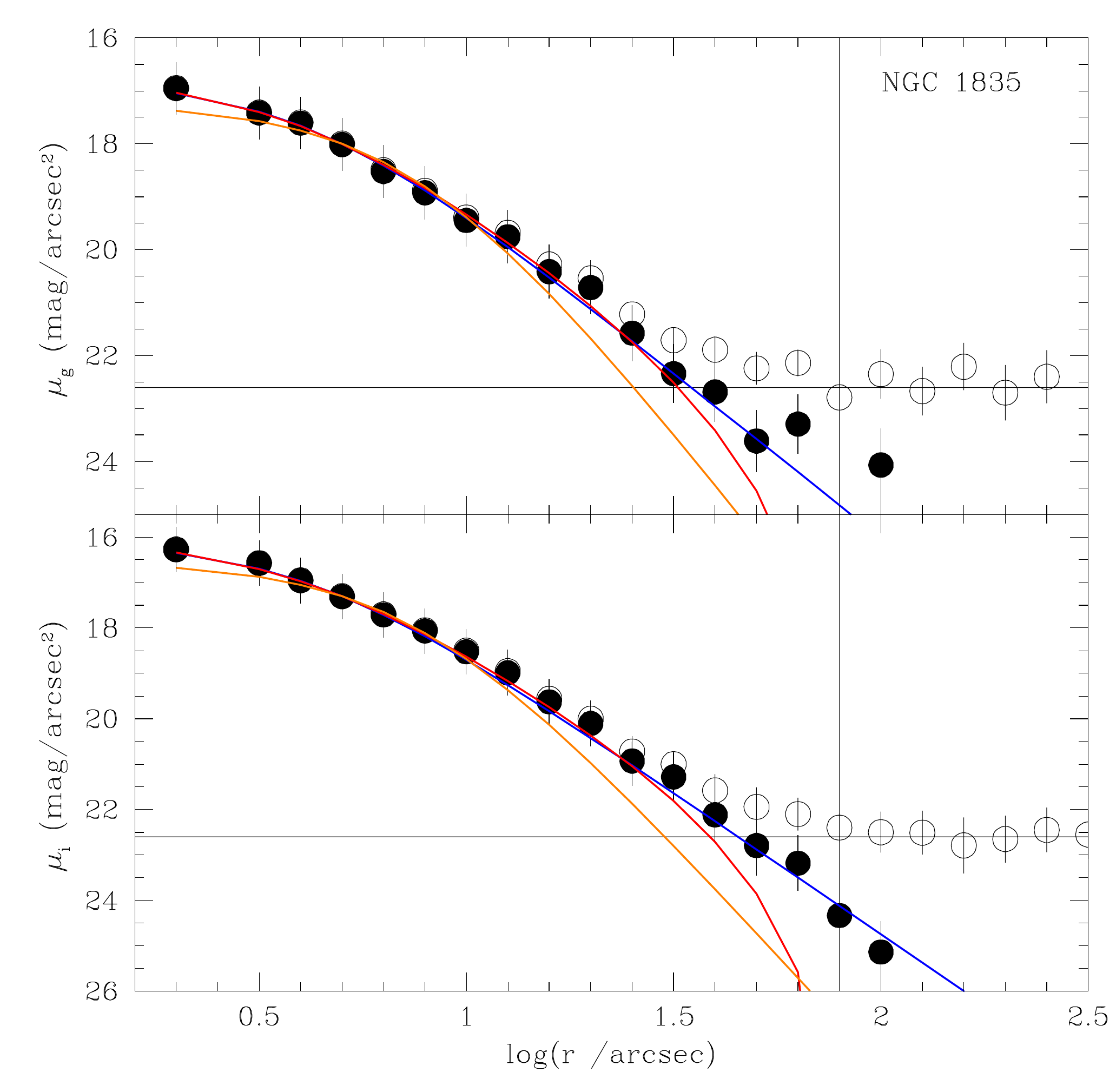}
    \caption{Same as Fig.~\ref{fig:fig3}, for NGC\,1835.}
   \label{fig:fig5}
\end{figure}

\begin{figure}
\includegraphics[width=\columnwidth]{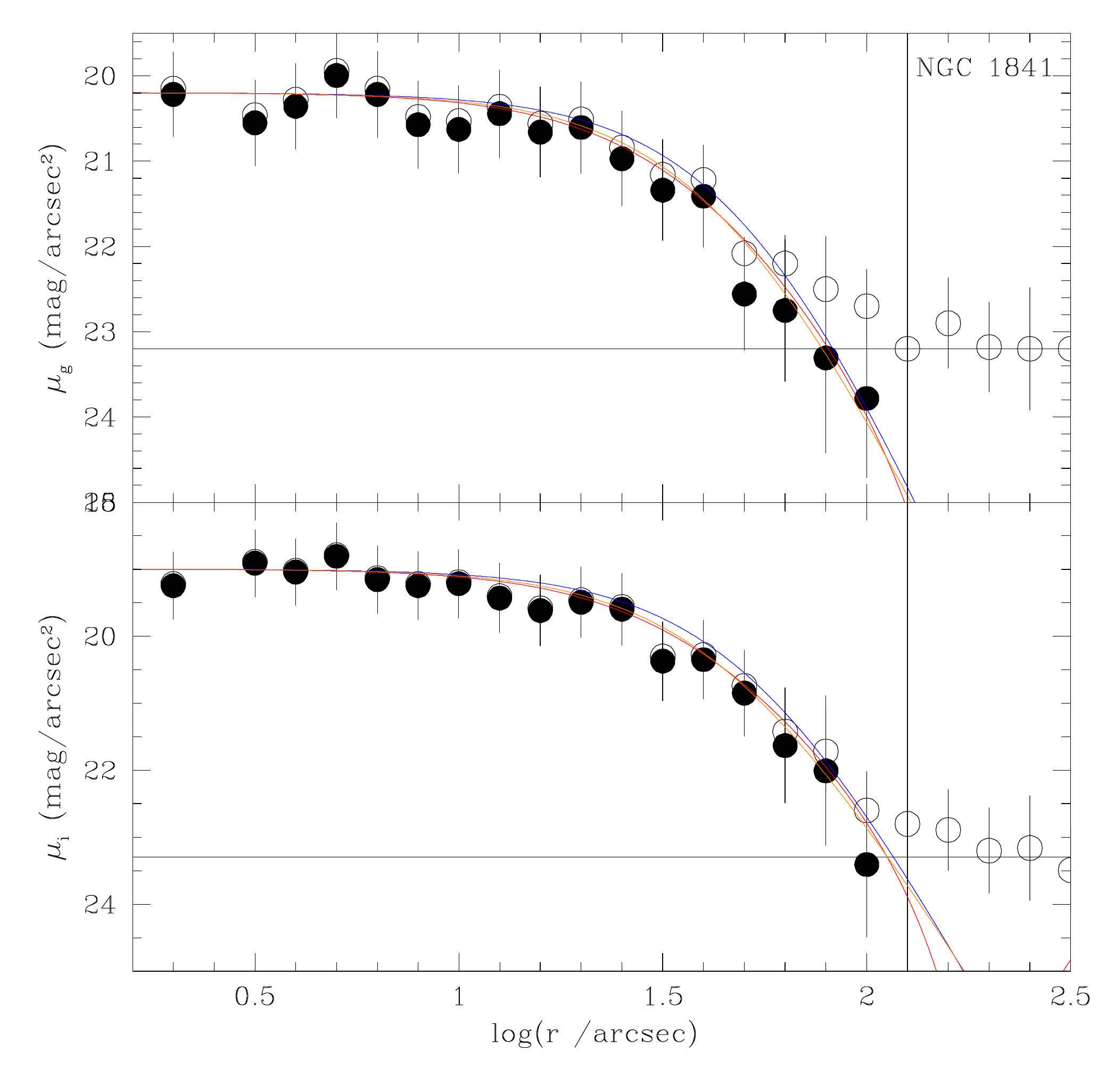}
    \caption{Same as Fig.~\ref{fig:fig3}, for NGC\,1841.}
   \label{fig:fig6}
\end{figure}

\begin{figure}
\includegraphics[width=\columnwidth]{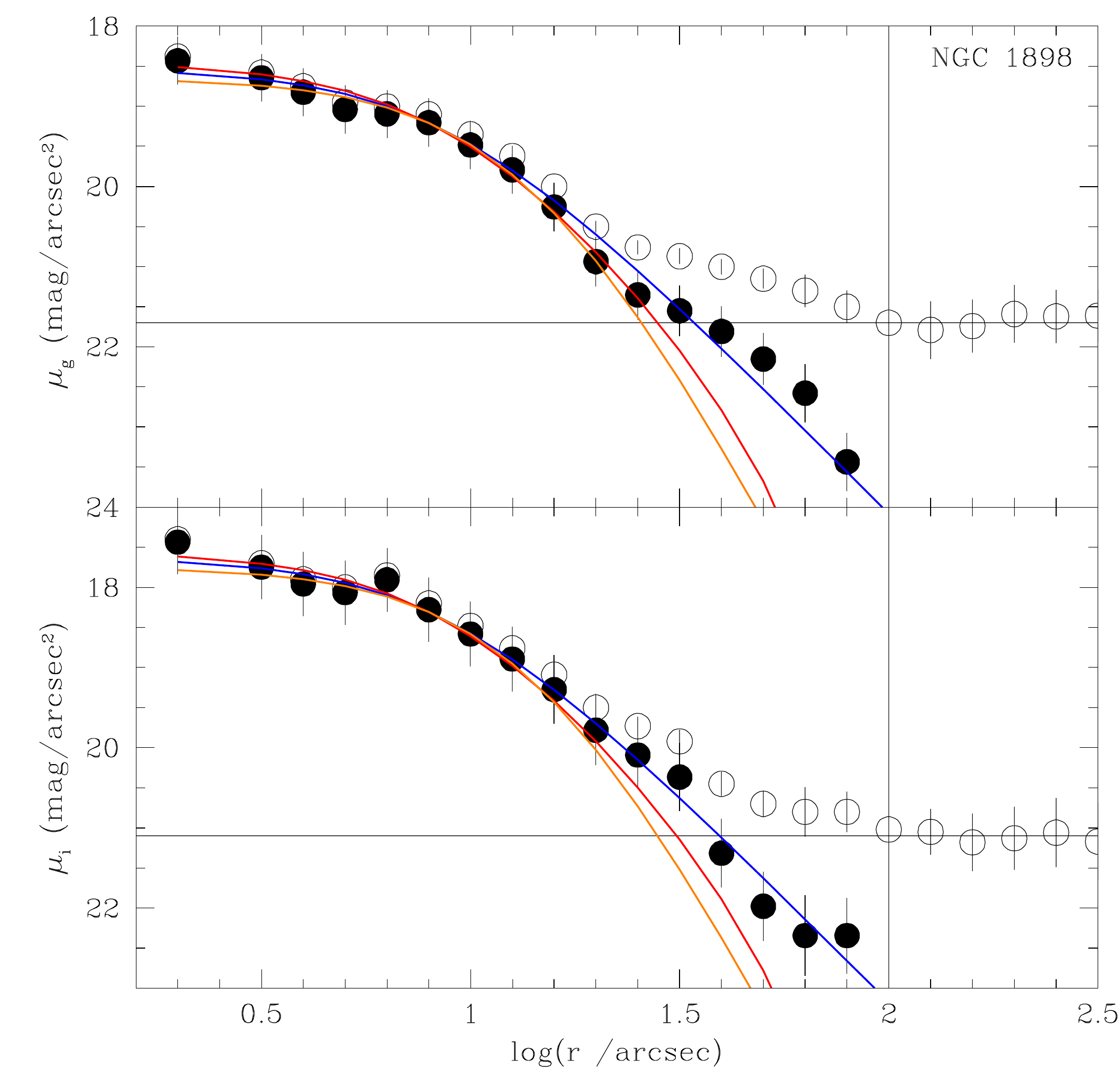}
    \caption{Same as Fig.~\ref{fig:fig3}, for NGC\,1898.}
   \label{fig:fig7}
\end{figure}

\begin{figure}
\includegraphics[width=\columnwidth]{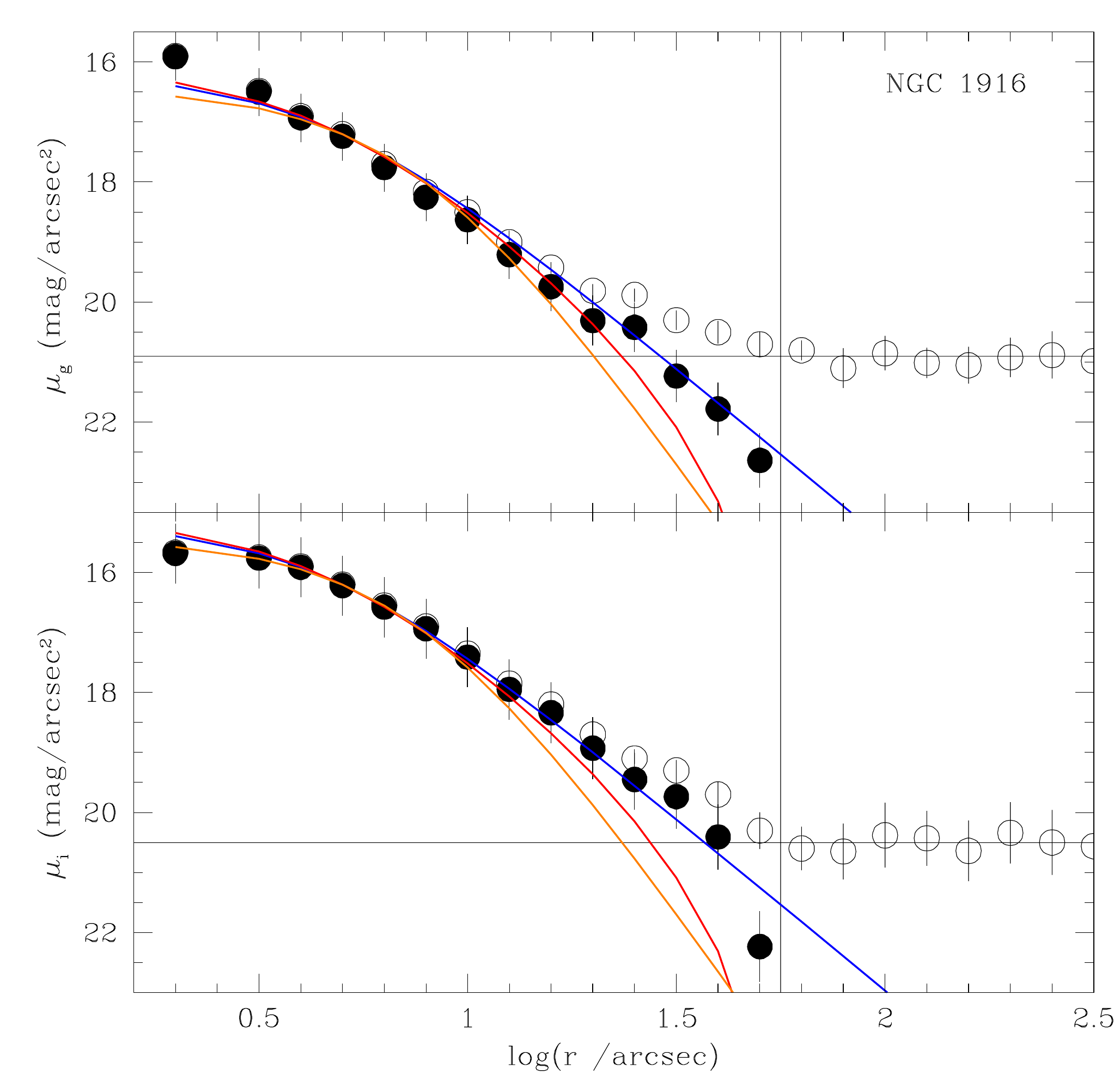}
    \caption{Same as Fig.~\ref{fig:fig3}, for NGC\,1916.}
   \label{fig:fig8}
\end{figure}

\begin{figure}
\includegraphics[width=\columnwidth]{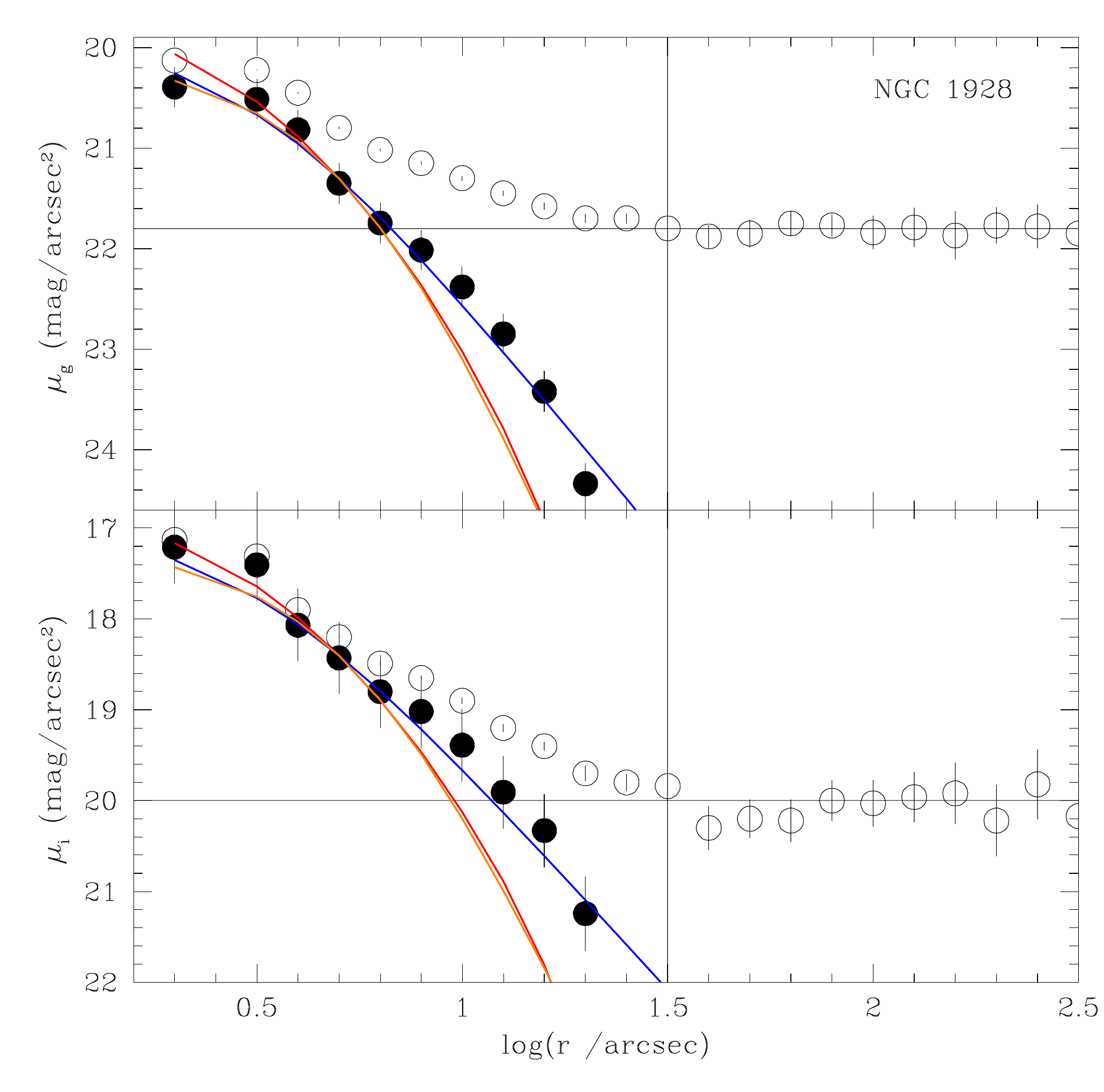}
    \caption{Same as Fig.~\ref{fig:fig3}, for NGC\,1928.}
   \label{fig:fig9}
\end{figure}

\begin{figure}
\includegraphics[width=\columnwidth]{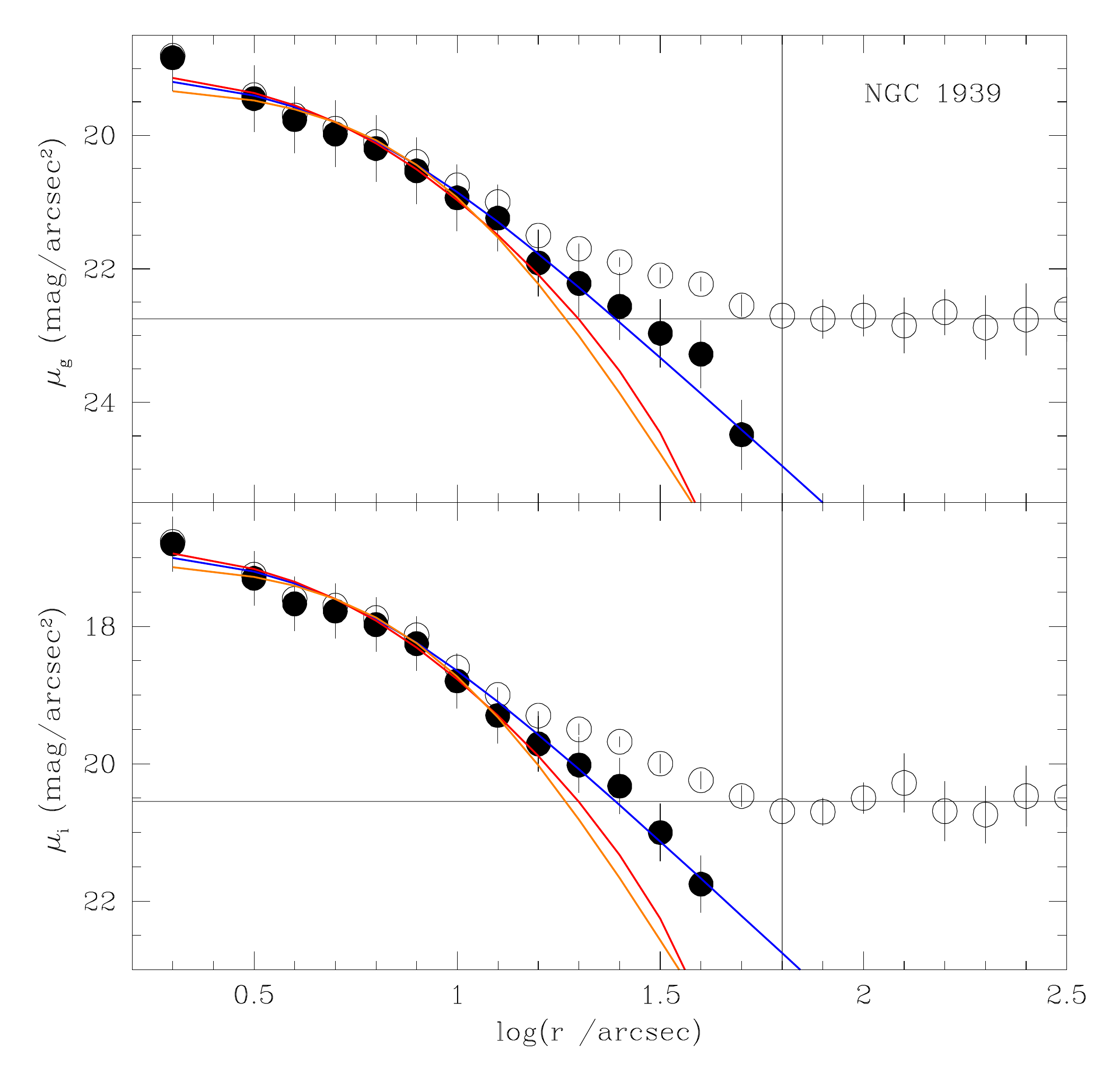}
    \caption{Same as Fig.~\ref{fig:fig3}, for NGC\,1939.}
   \label{fig:fig10}
\end{figure}

\begin{figure}
\includegraphics[width=\columnwidth]{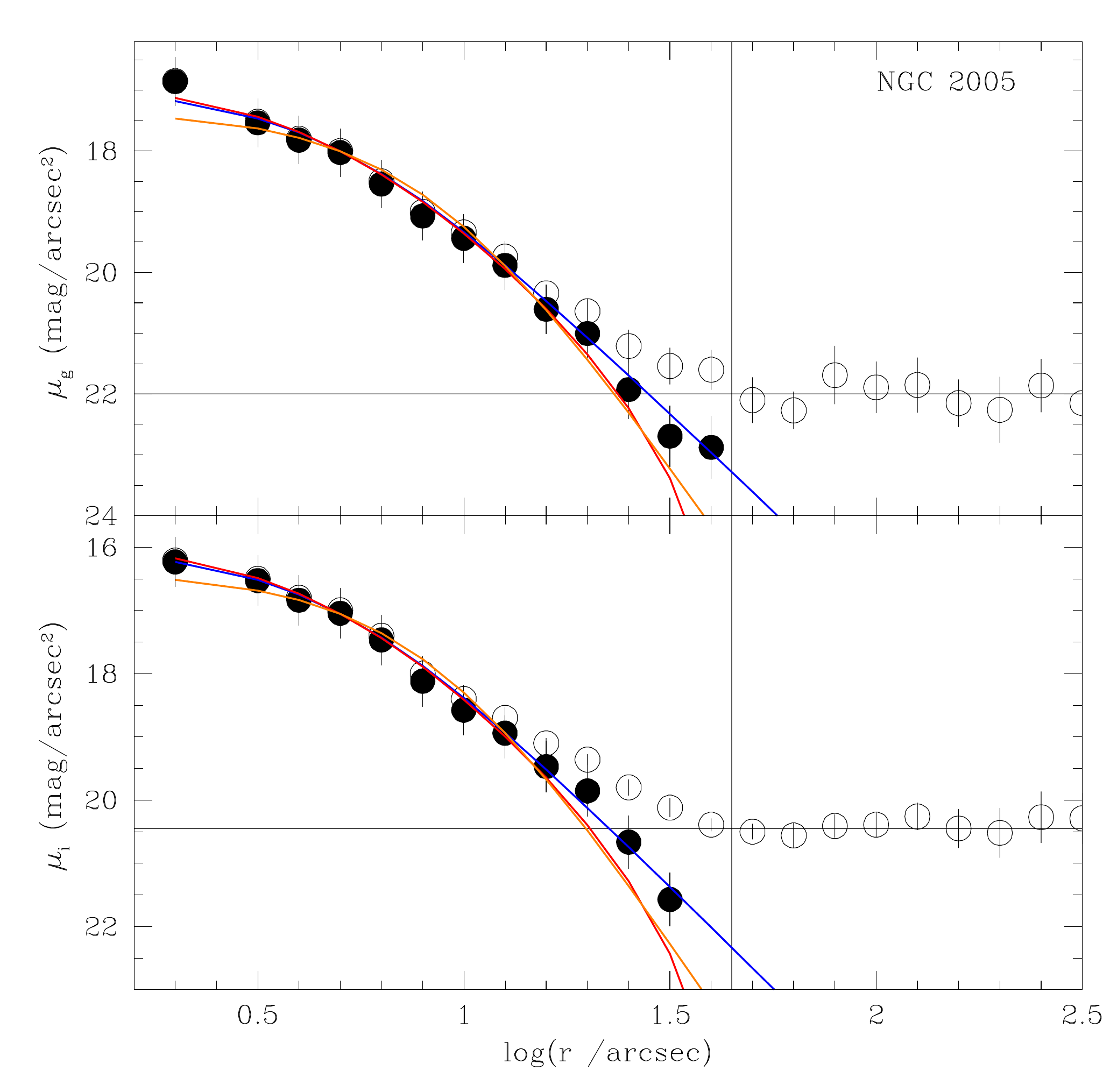}
    \caption{Same as Fig.~\ref{fig:fig3}, for NGC\,2005.}
   \label{fig:fig11}
\end{figure}

\begin{figure}
\includegraphics[width=\columnwidth]{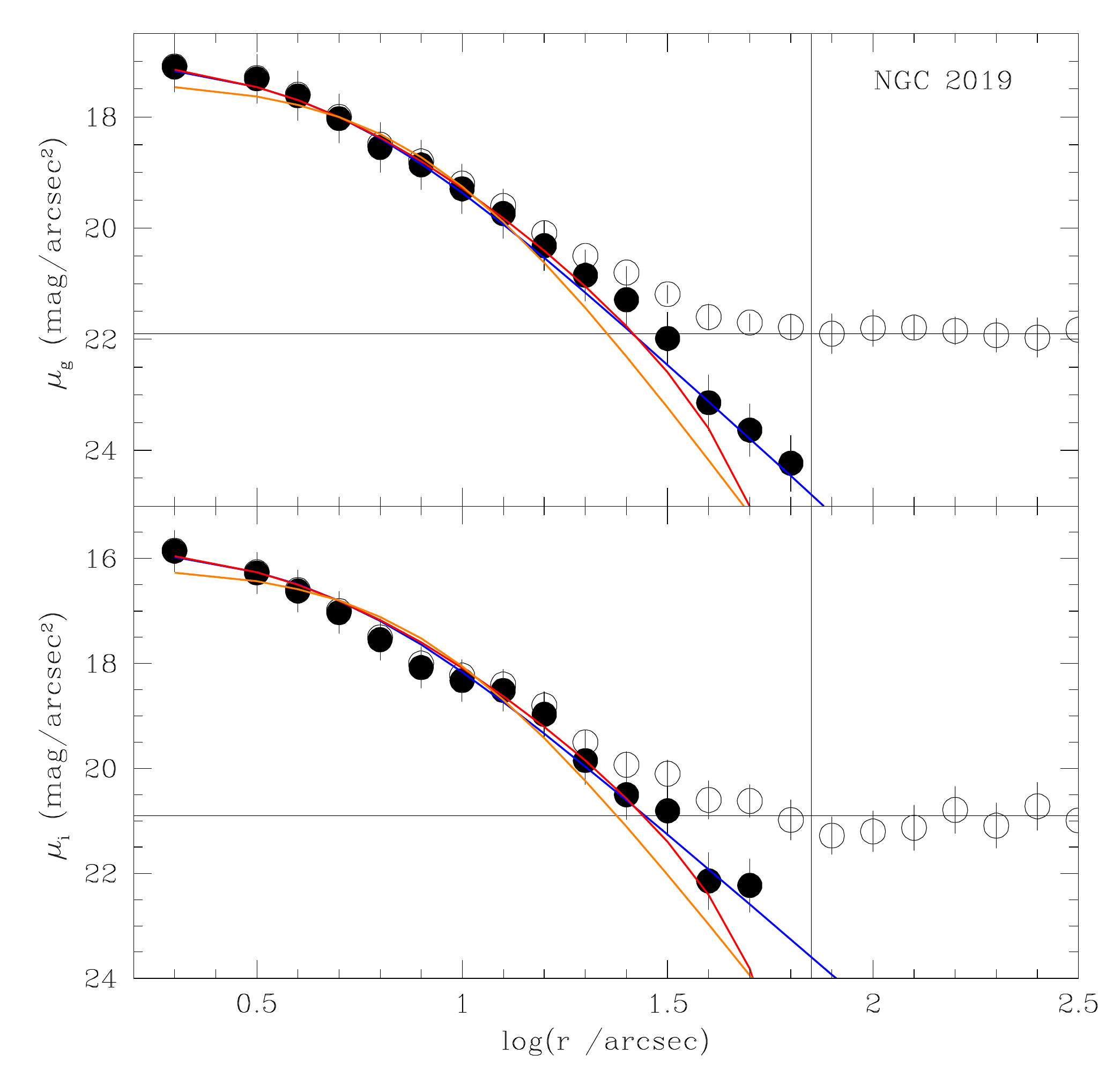}
    \caption{Same as Fig.~\ref{fig:fig3}, for NGC\,2019.}
   \label{fig:fig12}
\end{figure}

\begin{figure}
\includegraphics[width=\columnwidth]{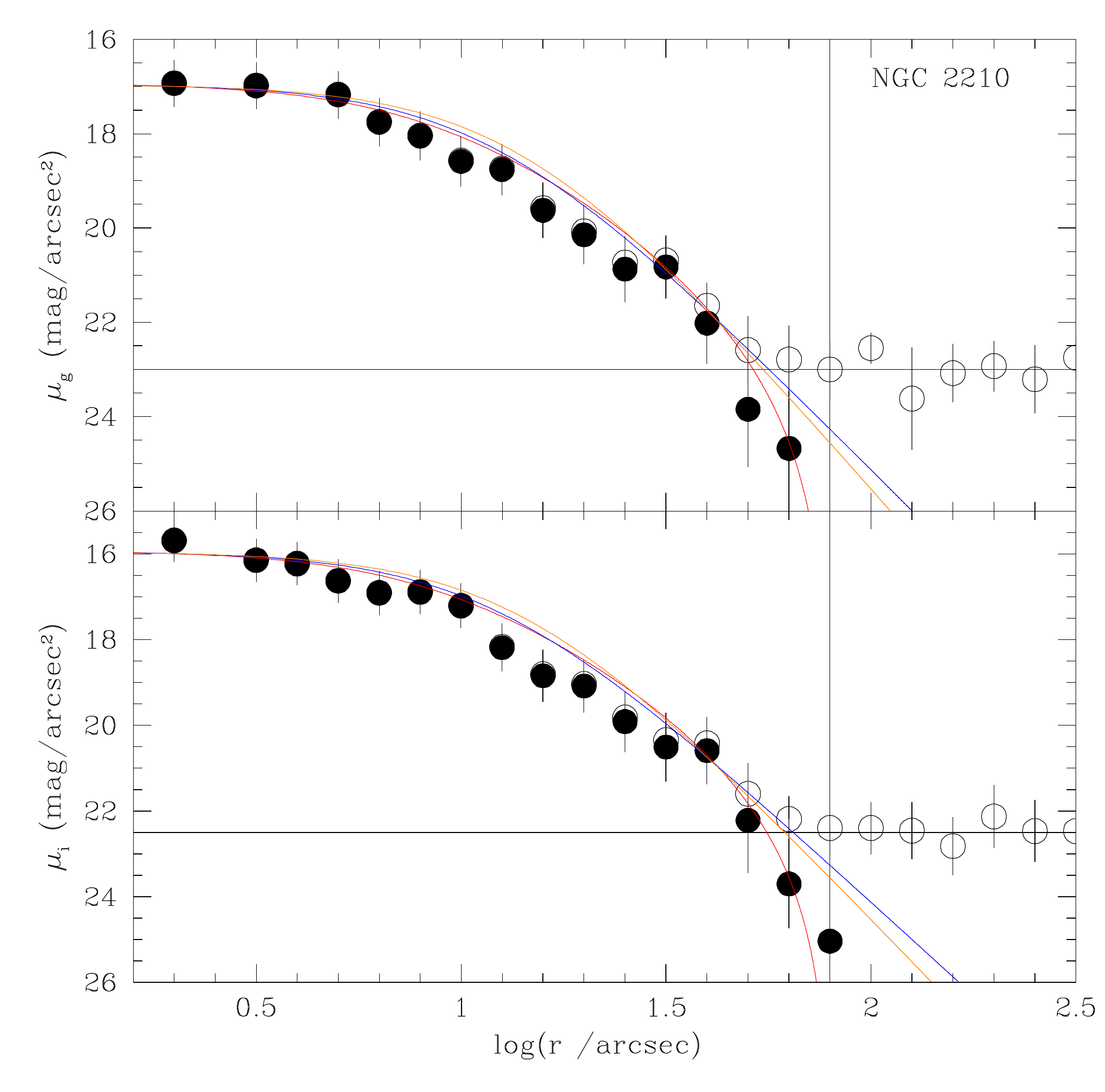}
    \caption{Same as Fig.~\ref{fig:fig3}, for NGC\,2210.}
   \label{fig:fig13}
\end{figure}

\begin{figure}
\includegraphics[width=\columnwidth]{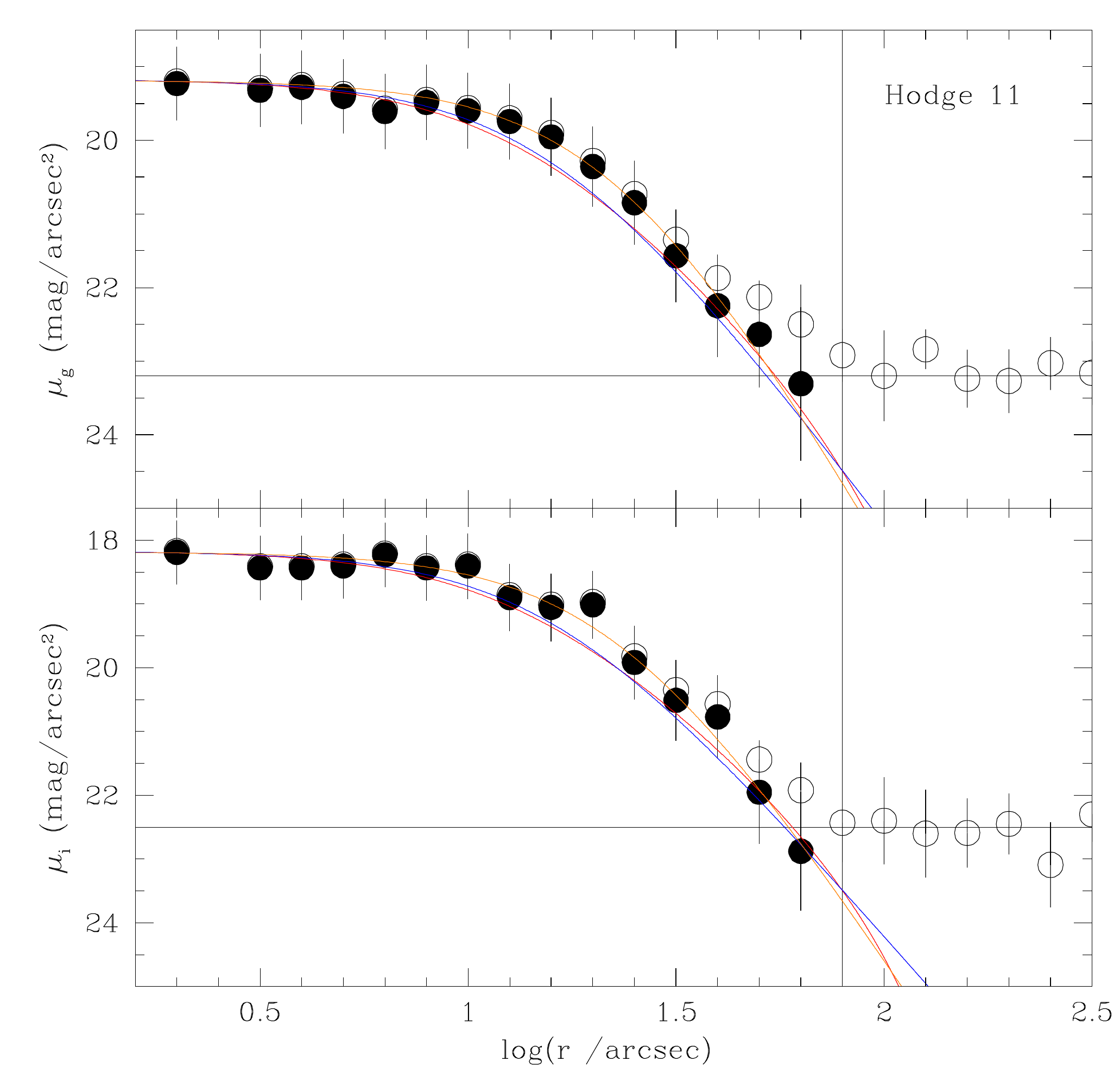}
    \caption{Same as Fig.~\ref{fig:fig3}, for Hodge\,11.}
   \label{fig:fig14}
\end{figure}

\begin{figure}
\includegraphics[width=\columnwidth]{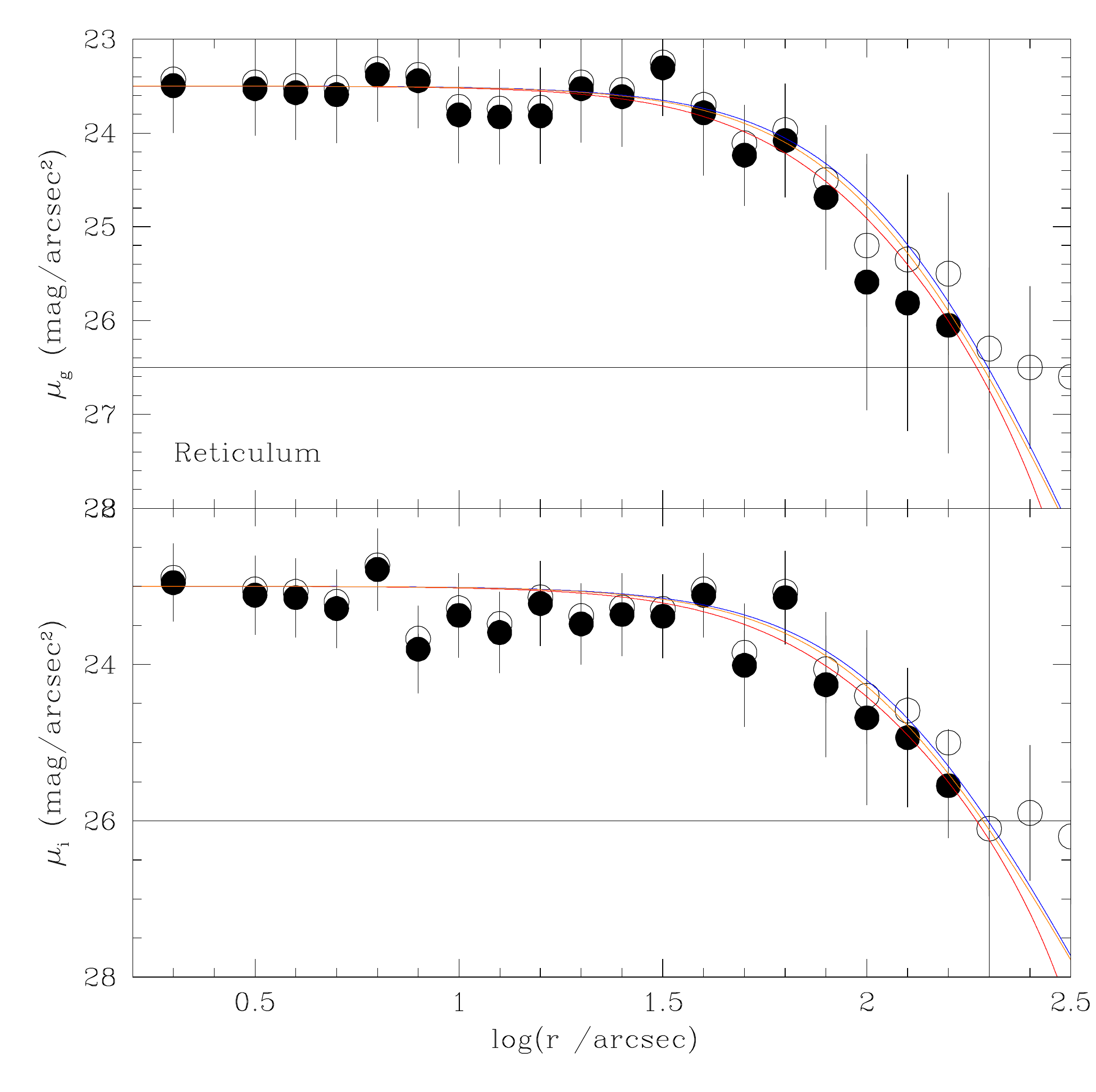}
    \caption{Same as Fig.~\ref{fig:fig3}, for Reticulum.}
   \label{fig:fig15}
\end{figure}

The resulting $g,i$ surface brightness radial profiles are depicted in 
Figs.~\ref{fig:fig3}-\ref{fig:fig15} with open circles. From them, we estimated the 
mean background levels by averaging those values for log(r /arcsec) $\ge$ 2.2
(horizontal line in the figures) that we subtracted from the
constructed surface brightness
profiles. The background region was chosen following the same criterion as in Section 3.1. 
Thus, we obtained the intrinsic cluster profiles that we used in the
subsequent analysis to estimate their structural parameters. They are represented
in Figs.~\ref{fig:fig3}-\ref{fig:fig15} with filled circles, while their errors were 
calculated by adding in quadrature both the uncertainties of the individual points 
and the mean error of the background level. We also estimated the cluster radii
($r_{cls}$) as the distance from the cluster centre where the mean background level 
intersects the cluster profile. These are represented by a vertical line in Figs.
\ref{fig:fig3}-\ref{fig:fig15}, while their values and uncertainties are shown
in Table~\ref{tab:table2}. As far as we are aware, this is the first time that
surface brightness (or stellar density) profiles are built for these GCs from
their centres out to $\sim$ 4 times their tidal radii.

We fitted EFF, \citet{king62} and \citet{plummer11}
models to the $g$,$i$ background subtracted surface brightness profiles in order
to derive $r_c$, $r_h$ and $r_t$ radii. 
While $r_c$ provides us with information about the innermost 
cluster structure, $r_t$ (King) and $\gamma$ (EFF) tell us about their outermost
regions. Notice that both King and EFF models depend on $r_c$, so that we used
two independent approaches to derive $r_c$.

We used a grid of $r_c$, $r_h$. $r_t$ and $\gamma$ values to fit independently
the three different models to the $g$ and $i$ surface brightness profiles, separately.
The values for which the models best resemble independently the $g$ and $i$
profiles were obtained by $\chi^2$ minimisation and then averaged.
These are shown in 
Table~\ref{tab:table2}, while Figs.~\ref{fig:fig3}-\ref{fig:fig15} depict the EFF, 
King and Plummer curves with blue, red and orange lines, respectively. 
For the resolved LMC GCs we superimposed the EFF, \citet{king62} and \citet{plummer11}
curves drawn in Fig.~\ref{fig:fig2}.
The present values of $r_c$ and $\gamma$ are in a general good agreement with those 
derived by \citet{mg2003} from surface density profiles using $HST$ F555W images. 
They did not obtain $r_t$ values because they barely reached $\sim$ 76 arcsec
out of the cluster centres due to the small HST field-of-view. For this reason, some of their $\gamma$ values - that
govern the outermost part of the profiles - could
also be systematically different. Additionally, they did not use the innermost points of the 
surface density profiles of some GCs (e.g. NGC\,1754, 1786, 1916)
that could explain the more different $r_c$ values.

From available GC ages \citep{piattietal2009,carrettaetal10} and masses $M_{\infty}$
\citep{mg2003}, we also computed half-mass relaxation times using eq. (2).
For the sake of the reader, the 
resulting $t_r$ values are listed in Table~\ref{tab:table2} along with the GC ages 
and masses. For completeness purposes, we finally included in Table~\ref{tab:table2} 
the deprojected GC distances to the LMC centre ($d_{deproj}$). 
We did not compute Jacobi radii for
clusters within 5 kpc from the LMC centre because, as far as we are aware, the variation of the
LMC mass with distance is not known for this inner distance range. Adopting a unique LMC mass
value for all the clusters lead to useless values.

\begin{table*}
%\centering
\caption{Derived properties of unresolved LMC GCs.}
\label{tab:table2}
\begin{tabular}{@{}lccccccccc}\hline
ID        &  $d_{deproj}$ & $r_c$  & $r_h$       &  $r_{cls}$     &  $r_t$         & $\gamma$ & $t_r$   & age$^a$   & log($M_{\infty}$ /$M_\odot$)$^b$ \\
          &  (kpc)        & (pc)   & (pc)        & (pc)           & (pc)           &          & (100$\times$Myr)   & (Gyr)  &                 \\\hline  
NGC\,1754 & 2.62 & 0.97$\pm$0.12 & 2.43$\pm$0.12 & 19.29$\pm$4.49 & 19.43$\pm$2.43 &2.4$\pm$0.3 & 4.4$\pm$0.6 & 13.0 & 5.39$\pm$0.30\\
NGC\,1786 & 2.27 & 0.85$\pm$0.12 & 2.43$\pm$0.12 & 19.29$\pm$4.49 & 19.43$\pm$2.43 &2.6$\pm$0.2 & 5.2$\pm$0.6 & 12.3 & 5.57$\pm$0.20\\
NGC\,1835 & 1.40 & 0.85$\pm$0.12 & 2.43$\pm$0.12 & 19.29$\pm$4.49 & 19.43$\pm$2.43 &2.5$\pm$0.2 & 6.6$\pm$0.5 & 13.4 & 5.83$\pm$0.13\\
NGC\,1898 & 0.43 & 2.07$\pm$0.12 & 4.62$\pm$0.24 & 24.29$\pm$5.64 & 24.29$\pm$2.43 &2.1$\pm$0.2 &18.5$\pm$2.8 & 13.5 & 5.88$\pm$0.30\\
NGC\,1916 & 0.17 & 0.97$\pm$0.12 & 2.43$\pm$0.12 & 13.65$\pm$3.13 & 14.57$\pm$2.43 &2.3$\pm$0.2 & 6.4$\pm$0.6 & 13.5 & 5.79$\pm$0.15\\
NGC\,1928 & 0.00 & 0.73$\pm$0.12 & 1.82$\pm$0.12 &  7.68$\pm$1.77 &  7.29$\pm$1.21 &2.0$\pm$0.2 & 1.7$\pm$0.2 & 12.3 & 4.87$\pm$0.20\\
NGC\,1939 & 0.43 & 1.21$\pm$0.12 & 2.92$\pm$0.12 & 15.33$\pm$3.55 & 14.57$\pm$2.43 &2.2$\pm$0.2 & 4.3$\pm$0.4 & 12.3 & 5.08$\pm$0.20\\
NGC\,2005 & 0.79 & 0.97$\pm$0.12 & 2.67$\pm$0.12 & 10.86$\pm$2.50 & 12.14$\pm$1.21 &2.6$\pm$0.2 & 5.5$\pm$0.5 & 13.5 & 5.49$\pm$0.16\\
NGC\,2019 & 1.05 & 0.97$\pm$0.12 & 2.67$\pm$0.12 & 17.20$\pm$3.96 & 17.00$\pm$2.43 &2.7$\pm$0.2 & 6.6$\pm$0.7 & 13.5 & 5.68$\pm$0.19\\
\hline
\end{tabular}

\noindent $^a$ taken from \citet{piattietal2009} and \citet{carrettaetal10}; $^b$ taken from 
\citet{mg2003}.
Note: to convert 1 arcsec to pc, we use the following expression,10$\times$10$^{(m-M)_o/5}$sin(1/3600)
where $(m-M)_o$ is the true distance modulus \citep[18.49 mag;][]{dgetal14}.
\end{table*}

\section{Analysis and discussion}

We started by considering the four GCs studied by WK17 (Hodge\,11, NGC\,1841, NGC\,2210 and
Reticulum) with the aim of exploring whether the
synchronicity between their ages and those of GGCs also reaches their structural parameters.
Fig.~\ref{fig:fig2} reveals that beyond the King's profiles none of the studied LMC GCs exhibit any sort of 
extra-tidal structures like those seen in a non-negligible number of GGCs \citep{carballobelloetal2012}.
Although the origin of such extended stellar structures around them is not well-known \citep{p17d}, some 
theoretical developments have shown that they could be due to potential escapers \citep{kupperetal2010} or 
potential observational biases \citep{bg2017}. According to \cite{go1997} and \citet{dinescuetal1999} 
some GGCs have experienced disruption by tidal shocks more important than 
by internal relaxation and evaporation, so that tidal tails should be
expected as debris from those interactions with the MW. Multiple tidal tails
from the interaction with the MW potential have also been recently predicted
from numerical simulations \citep{hk2015}. Therefore, in broad terms, we could conclude that
the GGC stars located in the outermost regions are
experiencing, in some way, gravitational effects due to the MW gravitational field.

Bearing in mind the above results, we added to our analysis the nine unresolved GCs
located inside a circle of 5 kpc from the LMC centre.
At first glance, every fitted curve (see Figs.~\ref{fig:fig3}-\ref{fig:fig15}) 
matches very similarly the two independent 
surface brightness profiles built per GC, which represent their present-day 
structural shapes as driven mainly by internal dynamical relaxation and LMC tidal 
forces. By inspecting Figs.\ref{fig:fig3}-\ref{fig:fig15} more closely, 
all the GCs do have outermost excesses of stars that are 
reproduced by the EFF models; neither King nor Plummer profiles are able to account 
for such a significant number of stars in the GC outskirts.
 This is a feature not seen in the outermost
LMC GCs studied above, whose extended stellar density radial profiles are 
satisfactorily reproduced by King's models. We speculate on the possibility that 
these observed stellar excesses might have reached the outskirts of the GCs under 
the effects of the LMC tidal field, since two-body relaxation would have led GCs to 
be simply tidally filled.

The orbital motions of LMC GCs are satisfactorily described by a  disc-like rotation with no GCs
appearing to have halo kinematics \citep{s92,getal06,shetal10}, so that it is expected that they do not cross 
the LMC disc as many GGCs do in the MW. Hence, they have not had chances to be subjected to tidal shocks which
could cause the appearance of extra-tidal structures, or in a more general context, that the LMC potential
has not been efficient in stripping stars off its GCs. The studied LMC GCs have $r_{cls}$ $\le$ $r_t$. 

\begin{figure*}
\includegraphics[width=\textwidth]{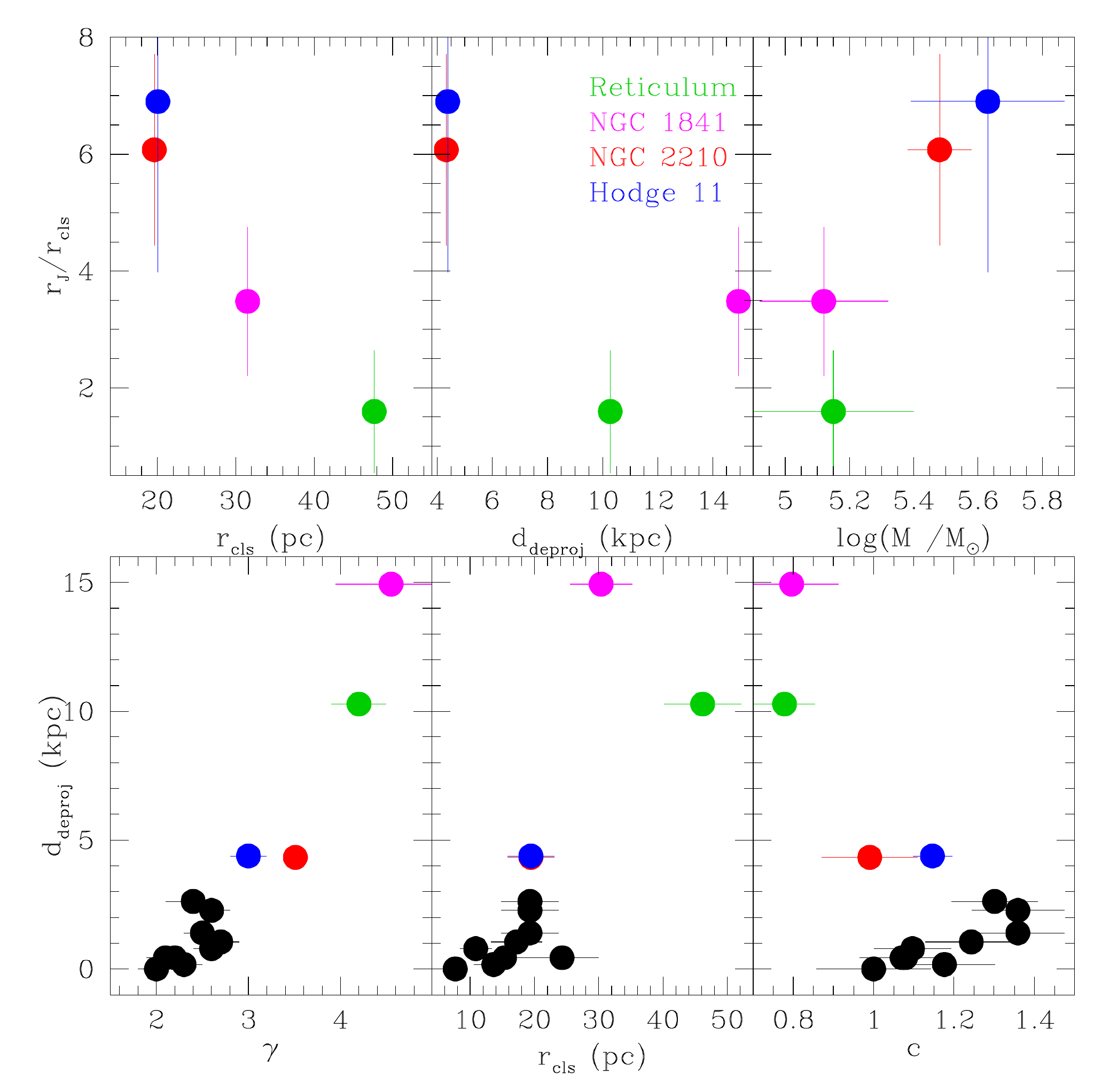}
    \caption{Relationships between different structural parameters of Reticulum (green), NGC\,1841 (magenta), 2210 (red), Hodge\,11 (blue) and unresolved LMC GCs (filled black).
}
\label{fig:fig16}
\end{figure*}

Since $r_J$ indicates 
where stars are gravitationally unbounded from the GC, we measured the degree of tidal filling by
comparing the derived $r_J$ values with the $r_{cls}$ ones (see top-left panel of Fig.~\ref{fig:fig16}).
As can be seen, GCs have not filled their respective Jacobi volumes, which explains the negative 
detection of extended stellar structures. Curiously, the more massive the GC the less filled the Jacobi volume, 
which, at a first glance, results contrary to the expected higher tidal filling in more massive GCs.
We found that such an apparent opposite behaviour could be due to the differential LMC gravitational field, i.e, GCs
located in the galaxy outskirts have been allowed to expand more than those in inner regions, 
provided that they were placed in disc-like orbits during their lifetimes \citep{miholicsetal2014}.
Precisely, the top-middle panel of Fig.~\ref{fig:fig16} (see also Table~\ref{tab:table1}) 
shows that Reticulum and NGC\,1841, located at $d_{deproj}$ of 10.62 and 15.55 kpc, respectively, are relatively
more expanded within their Jacobi volumes than NGC\,2210 and Hodge\,11 ($d_{deproj}$ $\approx$ 4.5 kpc), despite
the latter are more massive GCs. 

\begin{figure*}
\includegraphics[width=\textwidth]{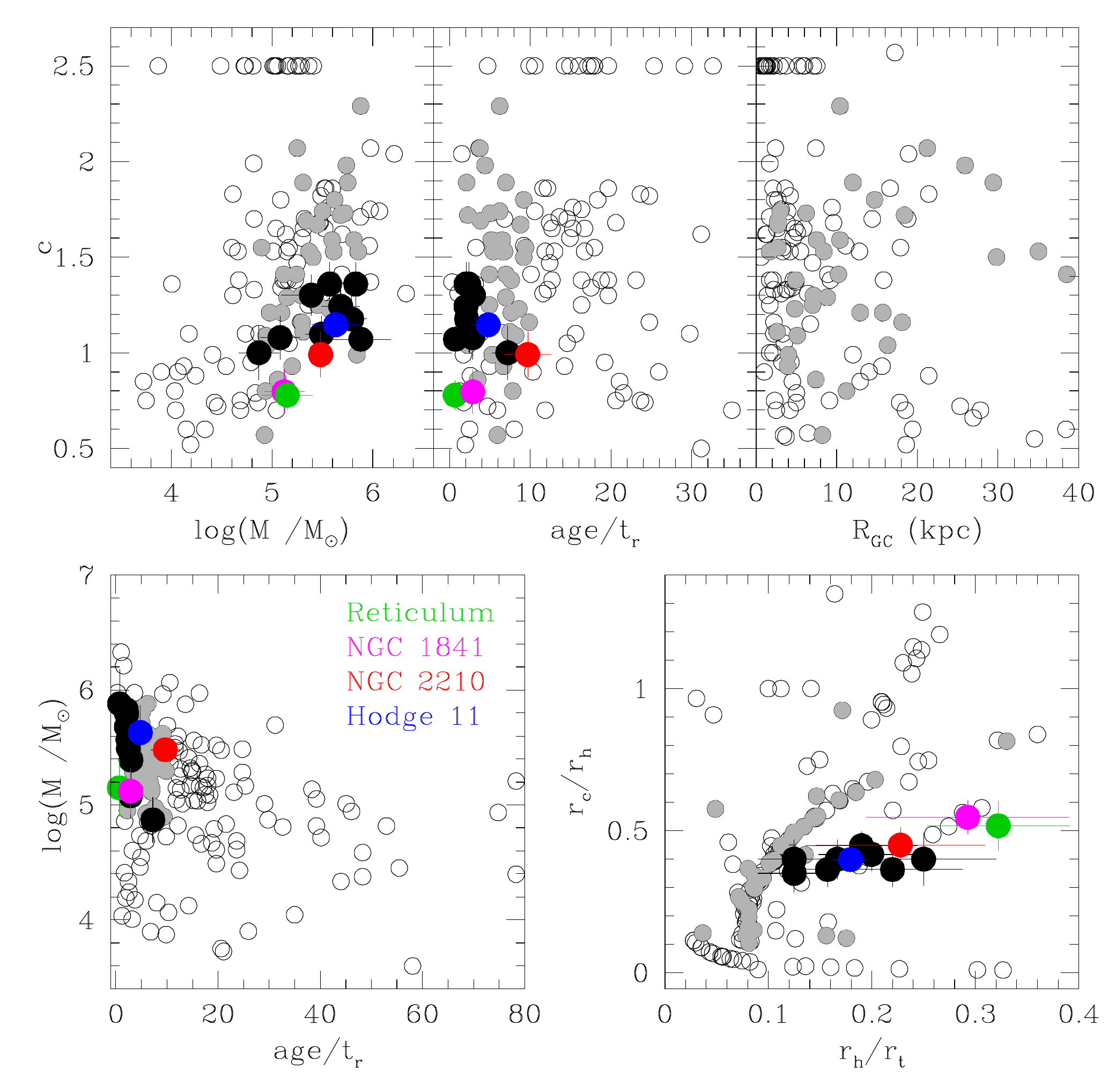}
    \caption{Comparison between different structural parameters of GGCs (open circles)
    with those of Reticulum (green), NGC\,1841 (magenta), 2210 (red), Hodge\,11 (blue)
and unresolved LMC GCs (filled black).
Grey filled circles represent GGCs with
    age/$t_r$ ratios and masses in the range of those for the LMC GCs. Note that
\citet{harris1996}  adopted $c$ = 2.5 for clusters believed to be core-collapsed.
}
  \label{fig:fig17}
\end{figure*}

This result interestingly suggests that, even though the LMC potential has been
relatively inefficient in carving extra-tidal structures, it has differentially affected the GC expansion due to
the internal dynamics (e.g., two-body relaxation). Indeed, the bottom panels of Fig.~\ref{fig:fig16}  would seem to suggest that $\gamma$ - the power-law index at large radii, i.e., the slope in a 
log-log plot -, the cluster radius ($r_{cls}$) and
the concentration parameter ($c$) show a trend with the position of the  GCs in the galaxy. These trends
are confirmed when the LMC GCs located within 5 kpc from the galaxy centre are included (see black circles
in Fig.~\ref{fig:fig16}). As can be seen, the smaller the $\gamma$ value, the stronger the LMC tidal field 
(see botom-left panel of Fig.~\ref{fig:fig16}),
suggesting that the  LMC gravitational field has been acting differently on the GCs.
Although the result could be somehow expected, it is now confirmed by the observations.
Similarly,  the innermost GCs are smaller than the outermost ones (see middle-panel of Fig.~\ref{fig:fig16}).
The variation in cluster size is not due to systematically different  masses:  GCs 
 with $d_{deproj}$ smaller or larger than $\sim$ 5 kpc clearly span comparable mass ranges
(see Tables~\ref{tab:table1} and \ref{tab:table2}).
Furthermore, even having similar masses, they have dynamically evolved differently: 
the innermost ones have lived for many more median relaxation times than their more 
remote counterparts (see bottom-left panel of Fig.~\ref{fig:fig17}).
This finding suggests that the LMC gravitational field has played an important role 
in accelerating the dynamical evolution of the innermost GCs. 

We compared the derived LMC GC structural parameters with those of GGCs using the values of $r_c$, $r_h$, $c$, $t_r$ 
and the Galactocentric distances $R_{GC}$ compiled by \citet[][December 2010 edition]{harris1996}. As for the GGC masses, we used masses
of 35 GGCs published by \citet{sb2017} and the integrated absolute magnitudes $M_V$ of \citet{harris1996} to
fit the expression log($M_{GGC}$ /M$_\odot$) =   -0.42$M_V$  + 2.02, which we used to estimate homogeneously the 
masses of the 157 GGCs included in the \citet{harris1996}' catalogue. Fig.~\ref{fig:fig17} depicts the results
of the comparison carried out for different relationships. 
GGCs are represented with open circles, while grey filled circles represent GGCs with age/$t_r$ ratios 
and masses similar to those of the LMC GCs (see bottom-left panel). Notice that, since \citet{harris1996} 
used the best measurements available in the literature to compile the GGCs' parameters, their errors
result smaller than the symbol size. Thus, we can straightforwardly draw conclusions about any connection
between these astrophysical properties in the GC populations of these  two galaxies.

The concentration parameter $c$ is a measure of the dynamical evolutionary stage of a cluster
following two-body relaxation, mass segregation and finally core-collapse
\citep{hh03}. Consequently, the larger the $c$ value
the more dynamically evolved a cluster. As can be seen in the top-left panel of Fig.~\ref{fig:fig17}, more
massive GGCs have dynamically evolved faster; while the LMC GCs 
have the lowest concentrations, thus implying that they are the least evolved, in 
comparison to their counterpart GGCs. This is an unexpected result because both GC 
populations have lived 
similar number of  times their $t_r$ (similar age/$t_r$ ratios), as is shown in the top-middle panel of Fig.~\ref{fig:fig17},
and are highly synchronized (WK17). Under the assumption that isolated GCs of similar masses should 
dynamically evolve similarly, the present outcome poses the idea that other conditions, for instance, the
gravitational potential of the host galaxy, the features of the GC orbital motions (halo- or disc- like orbits),
could also affect differentially the relationships between different structural parameters.

Indeed, in the $r_c/r_h$ vs $r_h/r_t$ plane (see see bottom-right panel of Fig.~\ref{fig:fig17}), where a cluster 
moves approximately in the top-right-bottom-left direction  \citep[][see, e.g., their figure 33.2]{hh03} while
relaxing toward a core-collapse stage, LMC GCs evolved differently than GGCs with similar masses and age/$t_r$ ratios. 
The figure shows that all the GCs are in the region corresponding 
to tidally filled star clusters, in
very good agreement with the above results. However,
the innermost GCs (black filled circles) have in average $r_{t}$ values relative to $r_h$
slightly larger than the outermost GCs (NGC\,1841, 2210, Hodge\,11 and Reticulum), because of the 
excess of stars in their outskirts that make the $r_h/r_t$ ratio smaller.
At the same time, the $r_c$/$r_h$ ratios of all the GCs are distributed
nearly in a narrow range, which suggests that there is no distinction of the 
dynamical evolution in the innermost regions of the LMC GCs.
Interestingly, there is a group of six GGCs
(NGC\,6352, 6749, 6838, Terzan\,4, 10 and 2MS-GC01) that follow approximately
the distribution of the LMC GCs. These GGCs have lived much more times their
relaxation times ($\Delta$(age/$t_r$) $\approx$ 15 - 50) and have
smaller masses ($\Delta$(log(M /M$_\odot$) $\approx$ 3.8 - 4.8) than the LMC GCs; all
of them are located at $R_{GC}$ $<$ 7 kpc.
On the other hand, $c$ values of GGCs do not seem to be linked with their positions 
in the MW (see top-right panel of Fig.~\ref{fig:fig17}), like LMC GCs appear to show respect to their own host 
galaxy (see bottom-right panel of Fig.~\ref{fig:fig16}). Nevertheless, some precaution in this suggested trend is needed, since
some GGCs change significantly their $R_{GC}$ because of their orbital motions, which could blur any
real relationship.

%\begin{figure*}
%\includegraphics[width=\textwidth]{fig18}
%    \caption{Relationships between different structural and dynamical
%    parameters. Black and red filled circles represent the nine GCs studied
%    here and those in Paper I, respectively. The black circles are clusters with 
%$d_{deproj}$ $<$ 5 kpc and the red circles are those with $d_{deproj}$ $>$ 5 kpc.}
%   \label{fig:fig18}
%\end{figure*}

\section{Conclusions}

We have made use of publicly available DECam images to build for the first time
stellar density and/or surface brightness radial profiles for almost all known old
LMC GCs. These are the most extended
stellar radial profiles built for these GCs so far.
The resulting radial profiles, which reach out to $\sim$ 4 times the GC tidal 
radii, allowed us to accurately estimate the respective mean background 
levels to be subtracted from the observed radial profiles.

We then matched the 
background subtracted radial profiles with EFF, King, and Plummer models. In doing this, we chose
core, half-mass and tidal radii which made the fitted models best resembled the traced 
radial
profiles. 
%Additionally, taking advantage of the known GC asymptotic masses, we estimated their Jacobi radii and  half-mass relaxation times. 
From the analysis of the relationships between different estimated dynamical 
and structural properties, we conclude:

$\bullet$ The studied GCs located farther than $\sim$ 5 kpc from the LMC centre would not seem to present any hint of extended stellar structures, like the family of
different characteristics (e.g., extra-tidal structures, tidal tails, extent halo-like features) seen in an important number of GGCs.  The apparent negative detection of such outer region structural tracers could suggest, contrarily to what
is predicted for some GGCs, that the LMC potential has not been efficient in stripping stars off 
its GCs. 

$\bullet$ Those GCs located closer than  $\sim$ 5 kpc from the LMC centre 
 show an excess of stars distributed in the outermost
regions with respect to the nominal amount of stars predicted by 
the best-fitting empirical King models. Such an excess of stars tightly depends on the position of the GCs in the galaxy, in such
a way that the closer the GC to the LMC centre, the larger the excess of stars. In addition,
the GC radii also show a remarkable trend with the position of the GC in the LMC disc,
namely: the farther the GC, the larger the GC radius. 

%Both outcomes are strongly confirmed 
%when the farthest GCs analysed in Paper I are added to the present sample. 

$\bullet$ These results can be interpreted in the light of their very well-known radial velocity  
disc-like kinematics. GCs rotate around the LMC centre keeping their mean distances 
from the LMC centre, so that they have mostly experienced the influence of the local 
gravitational field, which decreases as the distance from the LMC centre increases. 
Since the LMC tidal field is relatively more intense towards the inner parts of the galaxy, 
GCs placed
in orbits closer to the core of the galaxy have suffered relatively more severe effects,
for instance, that more GC stars moved outwards reaching the GC outskirts. Furthermore,
the gradual decrease of the LMC gravitational field with the distance from its centre
has also allowed GC to expand more as they occupy increasingly remote  positions in the galaxy.

$\bullet$ Moreover, we showed that outer LMC GCs ($d_{deproj}$ $>$ 5 kpc )  have not filled their respective Jacobi volumes and that
the more massive the GC the less filled the Jacobi volume. This result could be
%The latter is a striking outcome, that is
explained if their deprojected galactocentric distances are considered, because more distant GCs are
allowed to expand differentially more, provided that the LMC GCs have spent their lifetimes in disc- like
orbital motions.

$\bullet$ The different dimensions of the 13 analysed GCs do not seem to be related to
any systematic difference in their masses. Indeed, all the studied GCs have masses within the same mass range
independently of whether they are located inside or outside the 5 kpc circle. Despite this,
GCs with deprojected distances smaller than $\sim$ 5 kpc have lived 
for many more median relaxation times. This behaviour might also be a result of the stronger 
tidal fields, which have made the GCs dynamically evolved faster by bringing stars towards
the outermost GC regions sooner than what would have been expected in an scenario
of isolated two-body relaxation. However, we did not find any evidence that the GC regions 
inside their half-mass radii have been affected by the LMC tidal field.

%$\bullet$ The evolution of the LMC GC concentration parameter $c$ would seem to be dependent on both the internal dynamics and the position of the GC in the galaxy.

%$\bullet$ As compared with the GGCs that have lived a similar number of times their $t_r$ and have similar masses, the studied LMC GCs would seem to have the smallest $c$ values. 
%This finding would seem to suggest that other conditions, like the
%gravitational potential of the host galaxy or the features of the GC orbital motions (halo- or disc- like orbits),
%could play some role in the  the relationships between different structural parameters.

%$\bullet$ Such a different behaviour between the studied LMC GCs and GGCs is also seen in the  $r_c/r_h$ vs $r_h/r_t$ plane, and 
%in the fact that $c$ values of GGCs  do not seem to be correlated with their positions in the MW, although 
%due to their orbital motion some of them change their $R_{GC}$ significantly, which could hide any actual
%trend.

\section*{Acknowledgements}
This research draws upon data provided by David Nidever (SMASH, NOAO program ID 2013B-0440) as 
distributed by the Science Data Archive at NOAO. NOAO is operated by the Association of 
Universities for Research in Astronomy (AURA) under a cooperative agreement with the National 
Science Foundation. 

This project used public archival data from the Dark Energy Survey (DES). Funding for the 
DES Projects has been provided by the U.S. Department of Energy, the U.S. National Science
Foundation, the Ministry of Science and Education of Spain, the Science and Technology 
Facilities Council of the United Kingdom, the Higher Education Funding Council for England, 
the National Center for Supercomputing Applications at the University of Illinois at 
Urbana–Champaign, the Kavli Institute of Cosmological Physics at the University of Chicago, 
the Center for Cosmology and Astro-Particle Physics at the Ohio State University, the 
Mitchell Institute for Fundamental Physics and Astronomy at Texas A\&M University, 
Financiadora de Estudos e Projetos, Funda\c{c}\~{a}o Carlos Chagas Filho de Amparo \`{a}
Pesquisa do Estado do Rio de Janeiro, Conselho Nacional de Desenvolvimento Cient\'{\i}fico 
e Tecnol\'ogico and the Minist\'erio da Ci\`encia, Tecnologia e Inova\c{c}\~{a}o, the
Deutsche Forschungsgemeinschaft and the Collaborating Institutions in the Dark Energy 
Survey. The Collaborating Institutions are Argonne National Laboratory, the University 
of California at Santa Cruz, the University of Cambridge, Centro de Investigaciones 
Energ\'eticas, Medioambientales y Tecnol\'ogicas-Madrid, the University of Chicago, 
University College London, the DES-Brazil Consortium, the University of Edinburgh, 
the Eidgen\"{o}ssische Technische Hochschule (ETH) Zürich, Fermi National Accelerator 
Laboratory, the University of Illinois at Urbana-Champaign, the Institut de Ci\`{e}ncies
de l'Espai (IEEC/CSIC), the Institut de F\'{\i}sica d'Altes Energies, Lawrence Berkeley
National Laboratory, the Ludwig-Maximilians Universit\"{a}t M\"{u}nchen and the 
associated Excellence Cluster Universe, the University of Michigan, the National
Optical Astronomy Observatory, the University of Nottingham, the Ohio State University,
the University of Pennsylvania, the University of Portsmouth, SLAC National Accelerator
Laboratory, Stanford University, the University of Sussex, and Texas A\&M University. 
We thank the referee for his/her thorough reading of the manuscript and
timely suggestions to improve it. 

%%%%%%%%%%%%%%%%%%%%%%%%%%%%%%%%%%%%%%%%%%%%%%%%%%

%%%%%%%%%%%%%%%%%%%% REFERENCES %%%%%%%%%%%%%%%%%%

% The best way to enter references is to use BibTeX:

\bibliographystyle{mnras}
%\bibliography{paper} % if your bibtex file is called paper.bib

%to be uncommented before sending to editor
\input{paper.bbl}

% Alternatively you could enter them by hand, like this:
% This method is tedious and prone to error if you have lots of references
%\begin{thebibliography}{99}
%\bibitem[\protect\citeauthoryear{Author}{2012}]{Author2012}
%Author A.~N., 2013, Journal of Improbable Astronomy, 1, 1
%\bibitem[\protect\citeauthoryear{Others}{2013}]{Others2013}
%Others S., 2012, Journal of Interesting Stuff, 17, 198
%\end{thebibliography}

%%%%%%%%%%%%%%%%%%%%%%%%%%%%%%%%%%%%%%%%%%%%%%%%%%

%%%%%%%%%%%%%%%%% APPENDICES %%%%%%%%%%%%%%%%%%%%%

%\appendix

%If you want to present additional material which would interrupt the flow of the main paper,
%it can be placed in an Appendix which appears after the list of references.

%%%%%%%%%%%%%%%%%%%%%%%%%%%%%%%%%%%%%%%%%%%%%%%%%%

% Don't change these lines
\bsp	% typesetting comment
\label{lastpage}
\end{document}